\documentclass[a4paper,10pt,3p,twocolumn]{elsarticle}

\usepackage{graphicx}
\usepackage{hyperref}
\usepackage{siunitx}
\usepackage[version=4]{mhchem}
\usepackage{caption}
\usepackage{subcaption}
\usepackage{amsmath}

\journal{Astroparticle Physics}
\newcommand{\etal}{\textit{et al.\ }}

\begin{document}

\begin{frontmatter}

\title{Origins of Extragalactic Cosmic Ray Nuclei by Contracting Alignment Patterns induced in the Galactic Magnetic Field}

\author{M.~Erdmann}
\ead{erdmann@physik.rwth-aachen.de}
\author{L.~Geiger}
\author{D.~Schmidt}
\author{M.~Urban}
\author{M.~Wirtz}

\address{RWTH Aachen University, III. Physikalisches Institut A, Otto-Blumenthal-Str., 52056 Aachen, Germany}

\begin{abstract}
We present a novel approach to search for origins of ultra-high energy cosmic rays. These particles are likely nuclei that initiate extensive air showers in the Earth's atmosphere. In large-area observatories, the particle arrival directions are measured together with their energies and the atmospheric depth at which their showers maximize. The depths provide rough measures of the nuclear charges. In a simultaneous fit to all observed cosmic rays we use the galactic magnetic field as a mass spectrometer and adapt the nuclear charges such that their extragalactic arrival directions are concentrated in as few directions as possible. Using different simulated examples we show that, with the measurements on Earth, reconstruction of extragalactic source directions is possible. In particular, we show in an astrophysical scenario that source directions can be reconstructed even within a substantial isotropic background.
\end{abstract}


\end{frontmatter}

\section{Introduction}

In the past decade, research on ultra-high energy cosmic rays has advanced greatly. The key to progress are large-area observatories for extensive air showers equipped with modern detection techniques \cite{AbuZayyad:2012kk,ThePierreAuger:2015rma}. First, the arrival distribution of cosmic rays exhibits a significant departure ($>5\sigma$) from an isotropic distribution by a large-scale dipole structure \cite{Aab:2017tyv}. Second, the cosmic ray energy spectrum features a high-energy cut-off \cite{Abraham:2010mj,AbuZayyad:2013}. Third, the abundance of cosmic rays with larger nuclear masses increases with energy \cite{Aab2014a, Aab:2014aea}. The origin of cosmic rays, however, remains a burning research question. Although interesting candidate sources exist \cite{Abbasi2014,Aab:2018chp}, clear evidence for point sources are still to be found.

\begin{figure}[tt]
\begin{center}
\includegraphics[width=0.45\textwidth]{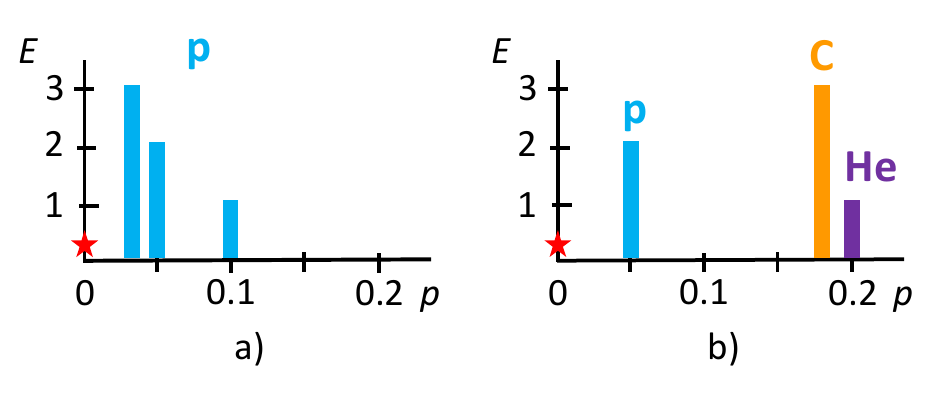}
\caption{Sketch of cosmic rays originating from the same source position (zero displacement, star symbol) with energy-dependent displacements $p$ for a)~protons, b)~a mixed composition.}
\label{fig:multiplets}
\end{center}
\end{figure}

Extrapolating from lower-energy cosmic ray measurements, ultra-high energy cosmic rays are likely protons or charged nuclei. Thus, deflections of cosmic rays in cosmic magnetic fields complicate searches for point sources by displacing charged particles from their original directions \cite{Stanev:1996qj, Harari2000, Harari:2002, Golup:2009, Giacinti:2010a, Golup:2011, Giacinti:2011}. For coherent magnetic fields, this seeming disadvantage is balanced by energy-ordered patterns in the arrival directions of e.g. protons which, in principle, allow cosmic rays from a single direction to be distinguished from an isotropic cosmic ray background. These alignment patterns are similar to the patterns typically found in spectrometers analyzing particle momenta (e.g. \cite{Abreu:2011md}). Fig. \ref{fig:multiplets}a shows a simplified example for protons at different energies. Note that prior knowledge of the magnetic field is not required here as the arrival directions and the energy arrangement together provide information on the direction and magnitude of the field. 

Most likely, the dominant deflections of cosmic rays appear within the galactic magnetic field \cite{Hackstein:2016pwa,Bray:2018ipq}. From numerous Faraday rotation and synchrotron emission measurements, parameterizations of the galactic field have been constructed featuring a highly inhomogeneous field with coherent and turbulent components \cite{Pshirkov2011, Jansson2012a, Pshirkov2013, Jansson2012b, Beck2014}. Thus, using the galactic field as a magnetic spectrometer is much more challenging compared to a laboratory experiment. Nevertheless, the field parameterizations enable detailed predictions for the deflection of cosmic particles as a function of their arrival directions, energy and charge \cite{Farrar:2014hma, Keivani:2014kua, emu2015, Farrar:2015dza, Farrar:2017lhm}. Analyses of the precision of the field parameterizations indicate that - for sufficiently large cosmic ray rigidities at least, i.e., energy divided by charge - parameterizations agree at a level sufficient for cosmic ray data analyses \cite{Erdmann:2016vle,Unger:2017kfh}.

To add to the challenge, a point source scenario of a cosmic ray accelerator with a mixed composition will change the energy arrangement in the arrival directions. Fig. \ref{fig:multiplets}b shows a simplified example of cosmic rays with the same energies as before. However, instead of protons (Fig. \ref{fig:multiplets}a), a mixed composition of a proton, a helium and a carbon nucleus is used, assigning the highest energy to the carbon ($Z=6$). In this case, the highest energy cosmic ray appears far away from the source as its deflection is largest owing to the large charge. A search method for origins of cosmic nuclei including corrections for such galactic magnetic field deflections has been presented in \cite{emuw2017}.

In order to exploit such mixed-composition alignment patterns induced by the galactic field, the cosmic ray charge needs to be estimated as well. Here, measurements of the air shower depth in the atmosphere provide at least a rough estimate of the cross section and thus of the nuclear mass, or the nuclear charge, respectively \cite{Aab:2014aea}.

In this paper, we present a novel method to decompose the distribution of cosmic ray arrival directions in terms of mixed-composition alignment patterns that arise from the galactic magnetic field. Using a global fit, we minimize the number of common cosmic ray directions outside the galactic magnetic field while simultaneously adjusting the cosmic ray charges. Optimization of fit parameters is performed through backpropagation technique known from training of neural networks.

This paper is structured as follows: First, we describe the basic analysis strategy for mixed-composition alignment patterns. Then we explain the fitting technique in detail. We present the main features of the method, initially using a $1$-dimensional scenario, before extending to the two dimensions on the surface of a sphere. Using a simulated astrophysical scenario, we show that point sources can be identified by correcting for the galactic magnetic field even in the presence of isotropic background. Finally, we present our conclusions.

\section{Basic strategy}

For each cosmic ray, the projection of the arrival direction outside the galaxy to its observed arrival direction on Earth is determined by the galactic field. Here, we rely on a field parameterization to reflect the true galactic field to a sufficient extent. 

Our basic idea is to fit a data set of observed cosmic rays to originate from similar directions outside the galaxy by varying the charges within their experimental uncertainties. The objective function of the fit minimizes the distances between cosmic ray arrival directions outside our galaxy and simultaneously constrains the measurements on Earth, namely the cosmic ray energy, arrival direction and composition. To obtain the significance of a measurement, the clustering strength obtained in the data is compared to scenarios of isotropic arrival directions. Alternatively, we use the final value of the objective function of the fit.

Initially, we explain the idea using a simplified $1$-dimensional scenario. In analogy to momentum measurements \makebox{$P\sim Ze\,B\, r$} of particles with a charge $Z e$ by the radius of curvature $r$ in a magnetic field $B$, we use a displacement from the original direction $s$ by energy- and charge-dependent translations to arrive at direction $p$:
\begin{equation}
p = s + \frac{Z}{E} 
\label{eq:translation}
\end{equation}
Here, the cosmic ray energy is denoted by $E$ and the charge by $Z$. The ratio $R=E/Z$ is usually referred to as rigidity. As a numerical example we allow for cosmic rays with energies between $E=1\hdots 10$ in arbitrary units, and charges between $Z=0\hdots 1$ (also in arbitrary units), such that the displacement is limited to between $0\le p-s \le 1$. 

\begin{figure}[tt]
\begin{center}
\includegraphics[width=0.4\textwidth]{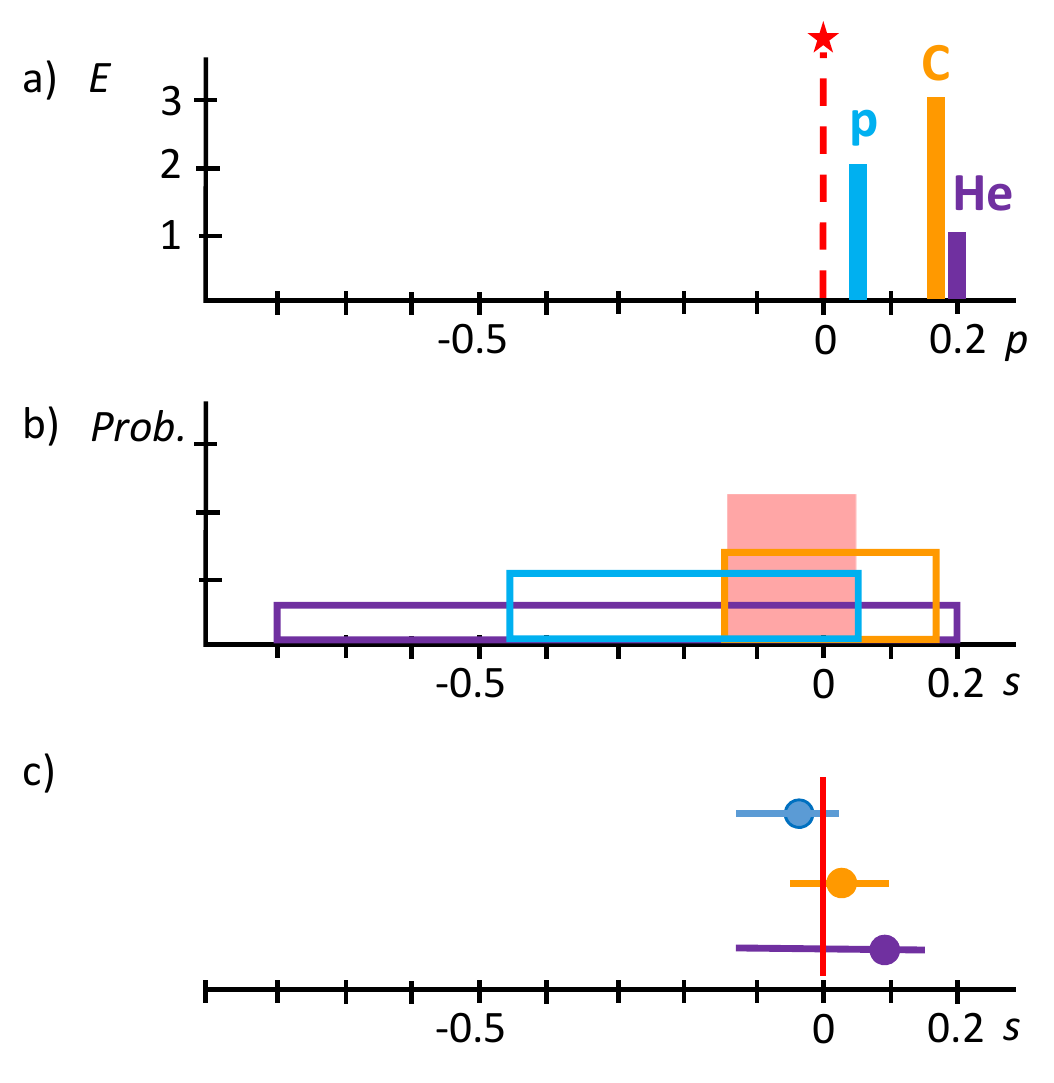}
\caption{Reconstruction ansatz of a common source from several cosmic rays for the mixed composition case, a)~true source with proton, helium and carbon nuclei, b)~overlap window from maximum possible charges and corresponding deflections, c)~weighted average source position from individual cosmic ray charge estimates. }
\label{fig:principle}
\end{center}
\end{figure}

In Fig.\ref{fig:principle}a, we show the mixed-composition alignment pattern of Fig.\ref{fig:multiplets}b together with the source position of the cosmic rays. In the two histograms below, we show different levels of potentially available information to reconstruct the true source position. 

If no information on the true cosmic ray charges is available, the range of possible source positions is limited by the maximum charge (Fig.\ref{fig:principle}b). For a low-energy cosmic ray of $E=1$, the displacement may be as great as $p-s=1$ if $Z_{max}=1$, while for cosmic rays with $E=3$ the maximal possible displacement is $p-s=1/3$. In the remaining interval the reconstructed source position follows a uniform probability distribution. Such estimates can be improved by using probability distributions obtained from measurements of the average shower depths instead of the intervals.

In Fig.\ref{fig:principle}c, we show the best possible case of having charge estimates for the individual cosmic rays available from measurements of the shower depth. Here, each cosmic ray provides an estimate of the reconstructed source position which can be improved by a weighted average of all estimated source positions.

Below, we will also study cosmic ray deflections on the 2-dimensional surface of a sphere. To this end, we will first analyze the effects of a horizontal magnetic field before investigating deflections in the galactic magnetic field.

\section{Techniques of the fit \label{sec:Aachen-fit}}

For each cosmic ray $i$ we observe the arrival direction $p_i$ and its energy $E_i$ on Earth. We also have one of the above-mentioned charge estimates $Z_i$ (compare Fig.~\ref{fig:principle}). 

A transformation prescription $T$ is used to predict for each given pair $(\hat{s}_i, \hat{Z}_i)$ of the estimated original cosmic ray direction $\hat{s}_i$ and the estimated cosmic ray charge $\hat{Z}_i$ the cosmic ray directions $\hat{p}_i$ on Earth:
\begin{equation}
\boxed{\hspace*{0.3cm}
\left(\begin{array}{c}
\hat{s}_i\\
\hat{Z}_i\\
\end{array}\right) \Longrightarrow T\left(\hat{s}_i, \hat{Z}_i, E_i \right) \Longrightarrow \hat{p}_i
\hspace*{0.3cm}}
\label{eq:fit-parameters}
\end{equation}

We optimize the directional parameters $\hat{s}_i$ and charge parameters $\hat{Z}_i$ according to the objective functions explained below. In doing so, we cluster the arrival directions $\hat{s}_i$ outside the galaxy while requiring consistency between predicted $\hat{p}_i$ and measured arrival directions $p_i$ as well as between estimated charges $\hat{Z}_i$ and measured shower depths $X_{\text{max}}$. For the fitting technique we use backpropagation as implemented in TensorFlow \cite{tensorflow}. For the optimization we use the concept of gradient descent.

\subsection{Transformations and parameters}

In the course of our study, we will increase the complexity of the scenario under investigation, and adapt the transformations and corresponding parameters accordingly.

\paragraph{One-dimensional transformations}
For our 1-dimensional studies, we use as the transformation $T$ the translation $T=\hat{s}+\hat{Z}/E$ presented in eq.~(\ref{eq:translation}). In the fit, the directional parameters $\hat{s}_i$ and $\hat{p}_i$ are scalar values, and the charge parameters $\hat{Z}_i$ are real numbers.

\paragraph{Two-dimensional transformations on a sphere}
On the surface of the sphere, the directional parameters $\hat{s}_i$ and $\hat{p}_i$ are 3-dimensional unit vectors with $(x,y,z)$-components, and the fitted charge parameters $\hat{Z}_i$ are real numbers.

As our transformation $T$ we use the rotation matrix $M(\delta)$ around the $z$-axis targeting perpendicular to the galactic plane, thereby varying only the longitude $l$ coordinate of the cosmic ray, while keeping its latitude $b$ constant. The rotation angles $\delta$ depend on the cosmic ray energies in units of EeV and their charges
\begin{equation}
\delta_i(\hat{Z}_i, E_i) = - 2\; \frac{\hat{Z}_i}{(E_i/\text{EeV})} \; .
\label{eq:angular_displacement}
\end{equation}
Here, the factor $2$ serves to displace heavy nuclei by several $10$ degrees. The transformations $T=M(\delta)$ in eq.~(\ref{eq:fit-parameters}) are then calculated according to
\begin{equation}
\hat{p}_i = M(\delta_i(\hat{Z}_i, E_i)) \cdot \hat{q}_i \;.
\label{eq:rotation}
\end{equation}

\paragraph{Deflections in the galactic magnetic field}
Finally, we use a representation $L$ of the galactic magnetic field to predict the arrival directions on Earth in eq.~(\ref{eq:fit-parameters}) with $T=L$ (see section~\ref{sec:GMF-sources}): 
\begin{equation}
\hat{p}_i = L(\hat{s}_i, \hat{Z}_i, E_i)
\label{eq:GMF}
\end{equation}
Obviously, this study is also performed on the surface of a sphere. The cosmic ray energies are in units of EeV, and their fitted charges are real numbers.

\subsection{Objective function \label{sec:objective}}

The objective function to guide the fit (\ref{eq:fit-parameters}) consists of several terms which aim at concentrating cosmic ray arrival directions prior to transformation while preserving the observed quantities, namely the arrival directions on Earth and the shower maximum.

\paragraph{Spatial distances between predicted and observed directions} The quality of the initial parameters, direction $\hat{s}_i$ and charge $\hat{Z}_i$ in eq.~(\ref{eq:fit-parameters}), can be judged by means of a comparison to the observed direction $p_i$. To this end, we define an objective term evaluating the quality of the initial parameters for all $N$ cosmic rays by:
\begin{equation}
D=\frac1N \sum_i \| p_i - \hat{p}_i \|^2
\label{eq:distance_loss}
\end{equation}

\paragraph{Differences between predicted and observed charges} To relate individual cosmic ray charge estimates $\hat{Z}_i$ to measurements of the shower depth of maximum $X_{\text{max}}$ we introduce an objective term which takes into account the asymmetry of the $X_{\text{max}}$-distribution that results from shower-to-shower fluctuations. This asymmetric distribution can be described by Gumbel functions $G(A_i, E_i)$ \cite{DeDomenico:2013wwa} where we assume $\hat{A}_i\approx 2 \hat{Z}_i$ for the atomic mass number. We construct the additional objective term $Q$ by demanding that the mean squared error of the $X_{\text{max}}$ values follow a $\chi^2_N$-distribution with $N$ degrees of freedom:
\begin{equation}
Q= \left[\frac1N \sum_i \frac{(X_{\text{max}, i} - \mu_i)^2}{\text{Var}(G(\hat{A}_i, E_i))} - 1\right]^2
\label{eq:charge_loss}
\end{equation}
Here, the sum is taken across all $N$ cosmic rays, $\mu_i = \operatorname*{arg\,max}_{X_{\text{max}}} G(\hat{A}_i, E_i)$ denotes the most probable $X_{\text{max}}$ value and $\text{Var}(G(\hat{A}_i, E_i))$ the Gaussian approximation of the variance for either the left or the right tail of the Gumbel distribution.

\paragraph{Clustering: spatial distances to common original directions} For the $1$-dimensional case, in order to demand the cosmic rays to originate from similar directions $\hat{s}_i$ we use the concept of $k$ nearest neighbors. The direction $\hat{s}_i$ of each cosmic ray should be close to the average direction $\langle \hat{s}_i \rangle$ of its $k$ nearest neighbors (knn). We calculate the average direction with:
\begin{equation}
\langle \hat{s}_i \rangle=\frac1k \sum_{\hat{s}_j \in \text{knn}(\hat{s}_i)} \hat{s}_j
\label{eq:center-nearest-neighbor}
\end{equation}

The objective term to demand clustering of original cosmic ray directions reads:
\begin{equation}
C=\frac1N \sum_i \|\hat{s}_i - \langle \hat{s}_i \rangle\|^2
\label{eq:cluster_loss}
\end{equation}

On the sphere we follow a slightly different approach. To force clustering, we apply a mean squared distance objective between all original cosmic ray directions of the form $\| \hat{s}_i - \hat{s}_j \|^2$. 

We only allow clustering of neighboring cosmic rays by additionally multiplying with an elliptical shaped weight factor $\epsilon_{ij}$ around each cosmic ray $i$, where the major axis is aligned with the elongated structures caused by the magnetic field. This particular shape reduces contributions from cosmic rays other than those originating from the same source. The elliptical weight factor reads $\epsilon_{ij} = \cos^{2\gamma}(\alpha_{ij})$, where $\alpha_{ij}$ denotes the angular distance between cosmic ray source estimations $\hat{s}_i$ and $\hat{s}_j$. We use $\gamma=4.3$ along the major axis to achieve a weight drop to $0.1$ at $\alpha_{ij}=40$~deg, and $\gamma=470$ along the minor axis for a weight drop to $0.1$ at $\alpha_{ij}=4$~deg. The complete cluster objective thus reads:
\begin{equation}
C = \frac{\sum_{\hat{s}_i} \sum_{\hat{s}_j} \epsilon_{ij} \cdot \|\hat{s}_i - \hat{s}_j \|^2}{\sum_{\hat{s}_i} \sum_{\hat{s}_j} \epsilon_{ij}}
\label{eq:center-pot-neighbor}
\end{equation}

\paragraph{Total objective function of the fit}
The fit is driven by an objective function consisting of the objective terms (\ref{eq:distance_loss}), (\ref{eq:charge_loss}), and depending on the study (\ref{eq:cluster_loss}) or (\ref{eq:center-pot-neighbor}). Their relative weights are adjusted by the hyperparameters $\lambda_Q$ and $\lambda_C$:
\begin{equation}
J = D + \lambda_Q Q + \lambda_C C
\label{eq:loss}
\end{equation}
As typical values we use $\lambda_Q=0.1$ and $\lambda_C=0.01$.

\section{Benchmark distributions of fits to mixed-composition patterns}

Initially, we benchmark the fit procedure using $1$-dimensional transformations as presented in eq.~(\ref{eq:translation}). We then extend to the $2$-dimensional sphere of the cosmic ray sky using similar linear transformations of the longitude while keeping the latitude constant (see eq.~(\ref{eq:rotation})). We use simulated mixed-composition data as will be explained below. 

\subsection{One-dimensional transformations}

As our data set for the $1$-dimensional case, we use pseudo cosmic rays with uniformly distributed energies between $E=1\hdots 10$ (arbitrary units) and charges with $Z=c/26$ for $c=1, \hdots, 26$. Estimating the nuclear mass by $A=2\,c$, we assign a value of the shower depth $X_{\text{max}}$ using the above-mentioned Gumbel functions $G({A}, E)$.

\begin{figure}[htt]
\captionsetup[subfigure]{aboveskip=-1pt,belowskip=-1pt}
\begin{centering}
\begin{subfigure}[b]{0.5\textwidth}
\includegraphics[width=\textwidth]{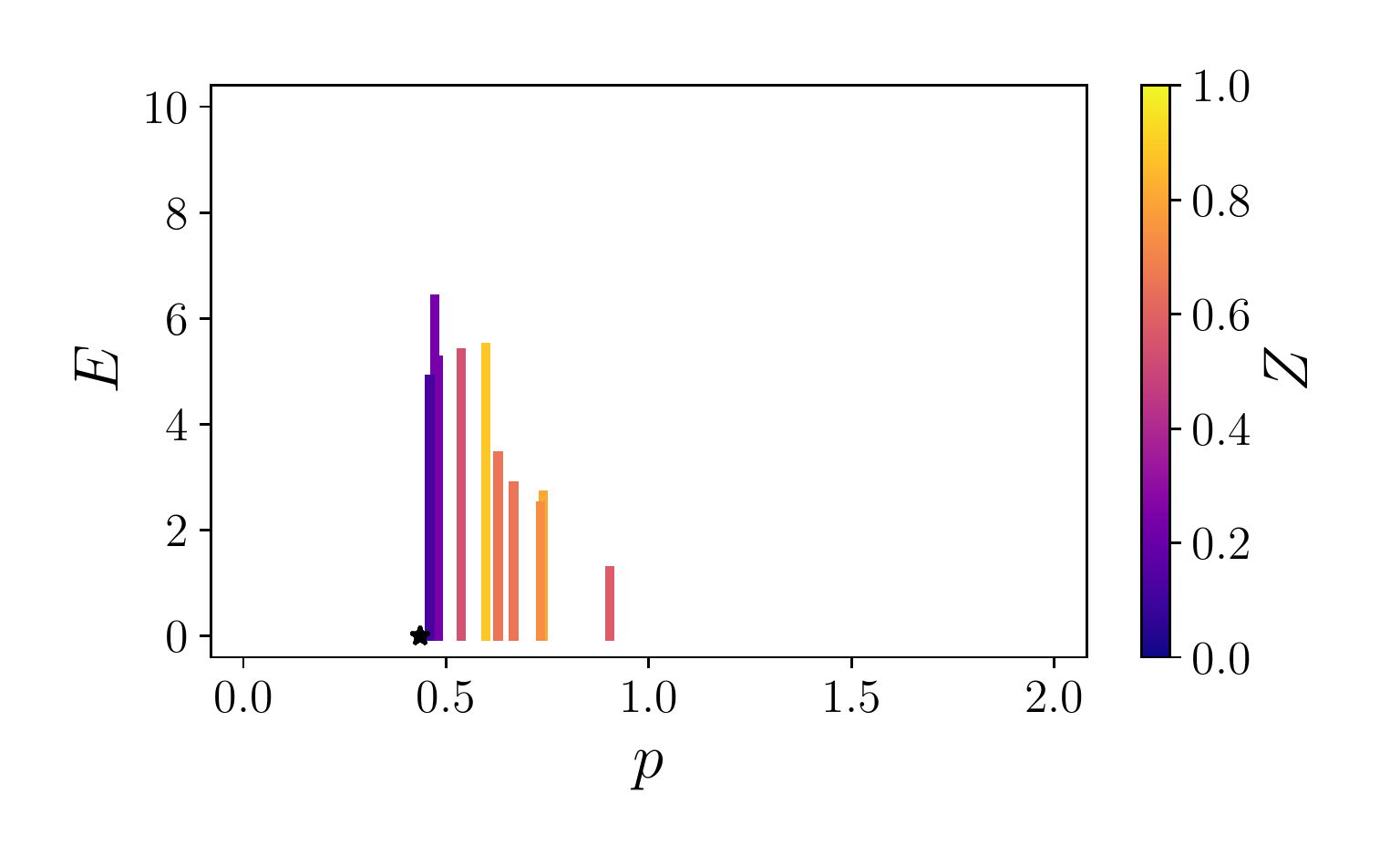}
\subcaption{}
\label{fig:1dscenario_a}
\end{subfigure}
\hspace{25mm}
\begin{subfigure}[b]{0.5\textwidth}
\includegraphics[width=\textwidth]{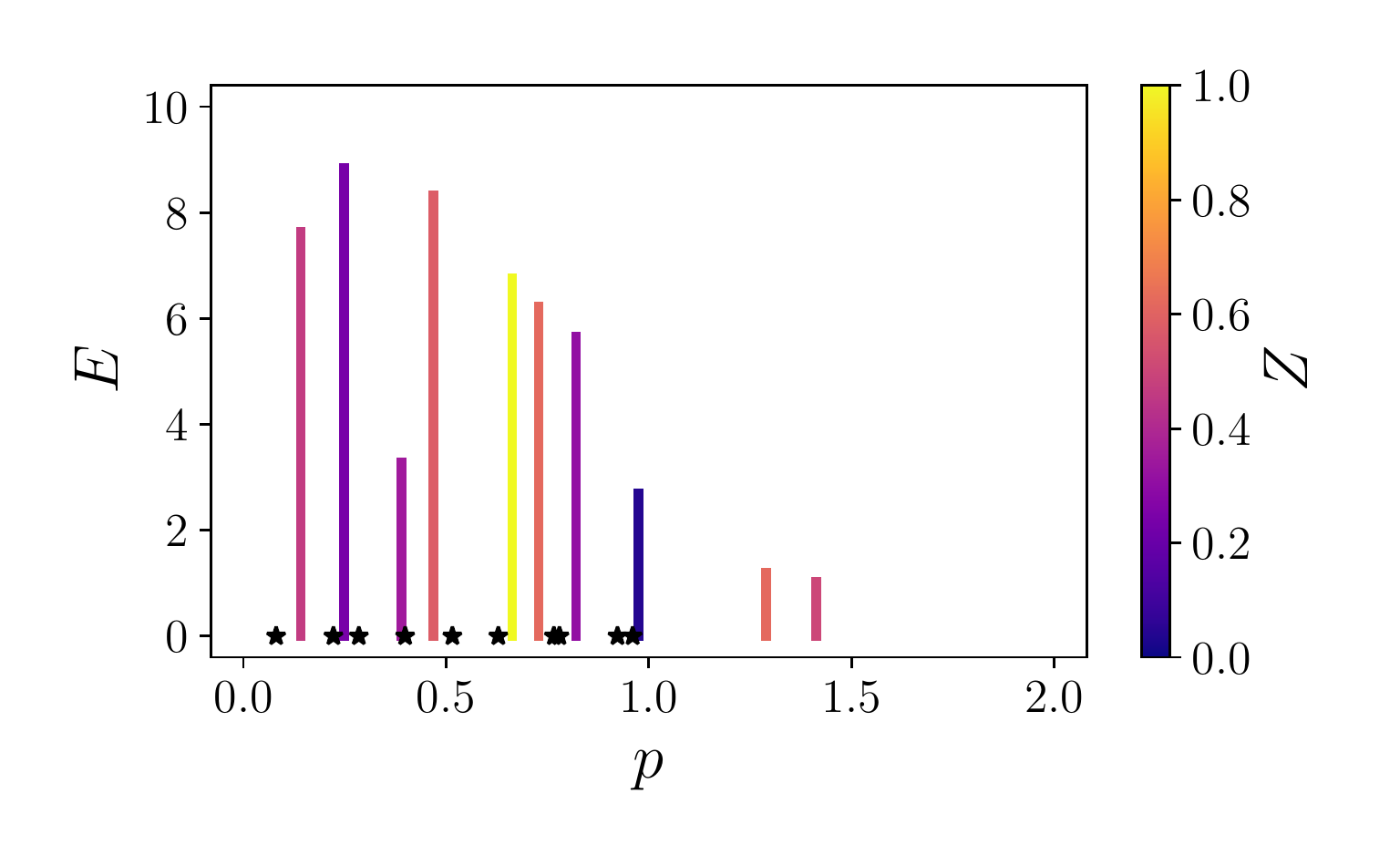}
\subcaption{}
\label{fig:1dscenario_b}
\end{subfigure}
\caption{Examples of $1$-dimensional displacements $Z/E$ in eq.~(\ref{eq:translation}) of cosmic rays (bars) from their source directions (black stars). The observed cosmic ray arrival direction is denoted by $p$, the cosmic ray energy is represented by the magnitude of the bar, and the charge by the color code. a)~Single source emitting $10$ cosmic rays, b)~isotropic scenario with $10$ sources each emitting $1$ cosmic ray.}
\label{fig:1dscenario}
\end{centering}
\end{figure}

In Fig.~\ref{fig:1dscenario_a} we show the direction of a single source using the star symbol, and its $N=10$ cosmic rays observed on Earth at directions $p_i$ using bars. The displacements were calculated using the transformation $T=Z_i/E_i$ in eq.~(\ref{eq:translation}). Cosmic ray energies $E_i$ are represented by the magnitudes of the bar symbols, and the true charges $Z_i$ are coded with the color scale. 

For the fit, we assign $\hat{Z}_i=0.5$ to the initial charge values in eq.~(\ref{eq:fit-parameters}) and calculate the initial original directions $\hat{s}_i$ by applying the inverse transformations $\hat{s}_i=p_i-\hat{Z}_i/E_i$ to the measured directions $p_i$.

\begin{figure}[htt]
\captionsetup[subfigure]{aboveskip=-1pt,belowskip=-1pt}
\begin{centering}
\begin{subfigure}[b]{0.475\textwidth}
\includegraphics[width=\textwidth]{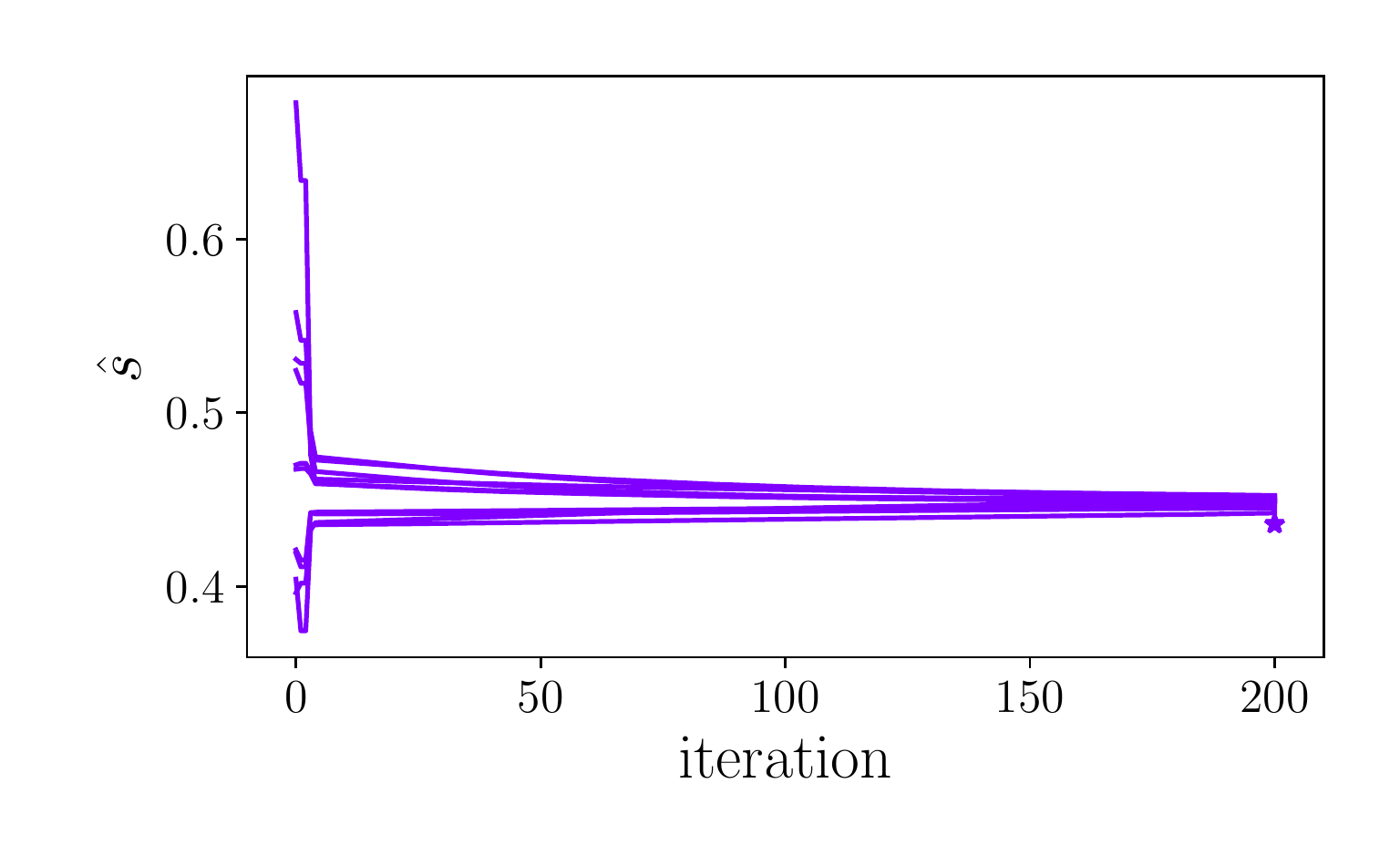}
\subcaption{}
\label{fig:1dsinglesource_a}
\end{subfigure}
\hspace{25mm}
\begin{subfigure}[b]{0.475\textwidth}
\includegraphics[width=\textwidth]{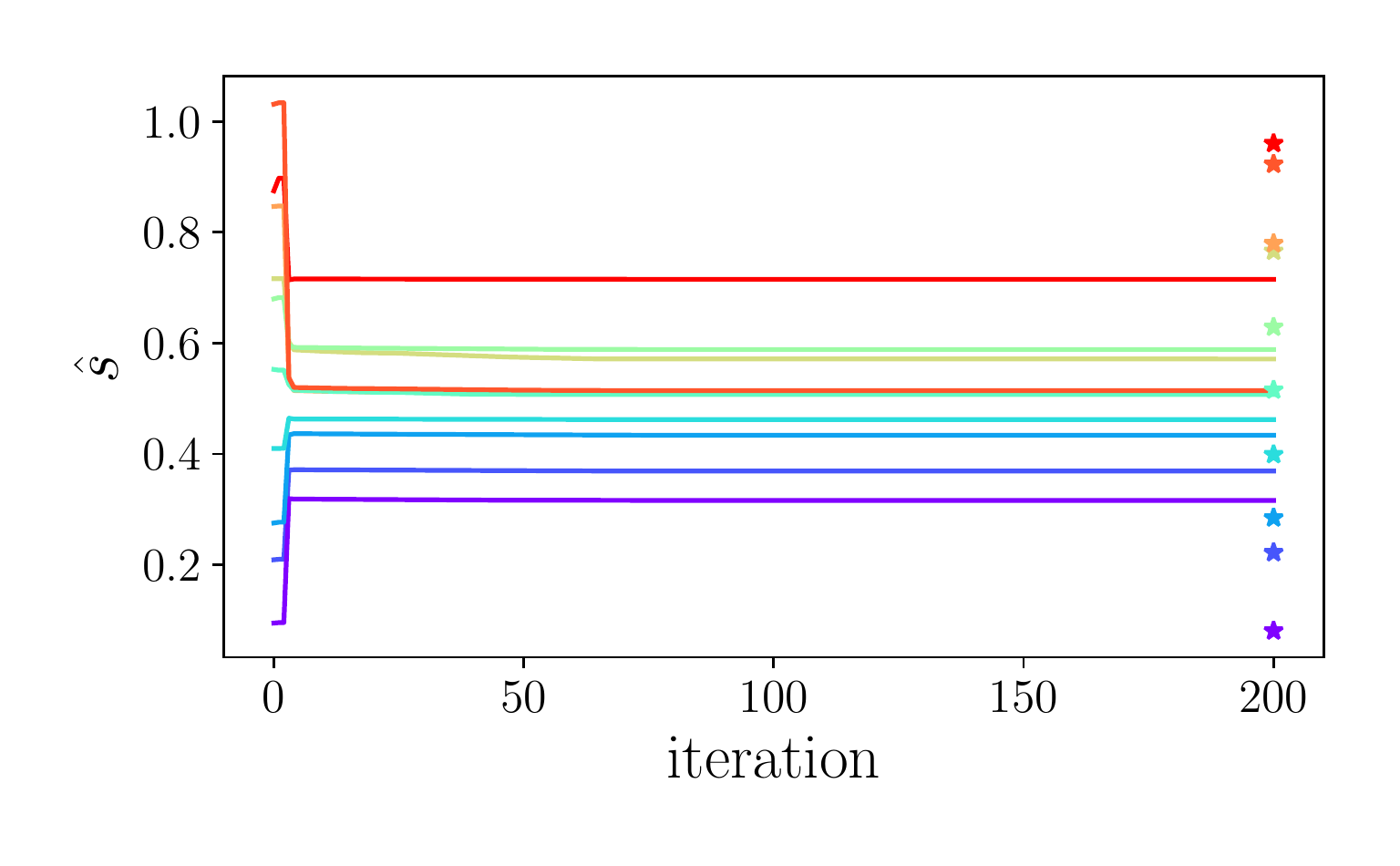}
\subcaption{}
\label{fig:1dsinglesource_b}
\end{subfigure}
\caption{For the examples of Fig.~\ref{fig:1dscenario}, we show for each cosmic ray the predicted direction $\hat{s}_i$ before displacement depending on the iterations of the fit. The true source directions are denoted by the stars. The fit varies the cosmic ray charges within the constraints of the measured arrival directions in (\ref{eq:distance_loss}) and shower depths (\ref{eq:charge_loss}) and attempts to concentrate the predicted directions in as few directions as possible (\ref{eq:cluster_loss}). a)~Single source with $10$ cosmic rays, b)~isotropic scenario.}
\label{fig:1dsinglesource}
\end{centering}
\end{figure}

In Fig.~\ref{fig:1dsinglesource_a}, we show for each cosmic ray the predicted direction $\hat{s}_i$ before transformation as a function of the number of fit iterations. The true source direction $s$ is shown by the star on the right. 

In each iteration of the fit, the average predicted direction $\langle \hat{s}_i\rangle$ of the $k$ nearest neighbors of cosmic ray $i$ is first calculated using eq.~(\ref{eq:center-nearest-neighbor}). In the example shown we use $k=N=10$. With the cluster term (\ref{eq:cluster_loss}) the fit is guided to modify each predicted direction $\hat{s}_i$ to move towards the average predicted direction $\langle \hat{s}_i\rangle$.

To modify the predicted direction $\hat{s}_i$ and still correctly predict the observed cosmic ray direction $p_i$ on Earth, as required in the distance term (\ref{eq:distance_loss}), the fit needs to adjust the cosmic ray charge $\hat{Z}_i$ accordingly. The charges are thus varied within the constraints of the measured shower depths $X_{\text{max}}$ using the charge term (\ref{eq:charge_loss}). In this way, consistency of the predicted source direction and adjusted cosmic ray charges with the measured arrival directions, energies, and shower depths is achieved.

To enable a comparison with the single-source scenario presented thus far, we also discuss scenarios of isotropic cosmic ray arrival. Isotropic arrival of $N=10$ cosmic rays can be interpreted as a scenario with $m=N$ sources each emitting a single cosmic ray. In Fig.~\ref{fig:1dscenario_b} we show as an example the directions of the $10$ sources using stars, and their cosmic rays observed on earth at directions $p_i$ using bars.

In Fig.~\ref{fig:1dsinglesource_b}, we show for each cosmic ray the predicted direction $\hat{s}_i$ before transformation depending on the fit iterations. The true source directions are shown using stars on the right, where the color code relates each cosmic ray with its respective source. 

Again, for each cosmic ray $i$ the fit averages the predicted direction $\langle \hat{s}_i\rangle$ of the $k=10$ nearest neighbors (\ref{eq:center-nearest-neighbor}), and attempts to concentrate the predicted directions according to the cluster term (\ref{eq:cluster_loss}). However, to modify the predicted direction $\hat{s}_i$, the observed cosmic ray direction $p_i$ on Earth is to be preserved in the distance term (\ref{eq:distance_loss}) which requires an adjustment of the cosmic ray charge $\hat{Z}_i$ according to the charge term (\ref{eq:charge_loss}). 

In the balance of the different objective terms (\ref{eq:loss}), the fit is driven to split the predicted directions into several clusters with low cosmic ray occupancy as can be seen in Fig.~\ref{fig:1dsinglesource_b}. 

\begin{figure}[ttt]
\captionsetup[subfigure]{aboveskip=-1pt,belowskip=-1pt}
\begin{centering}
\begin{subfigure}[b]{0.475\textwidth}
\includegraphics[width=\textwidth]{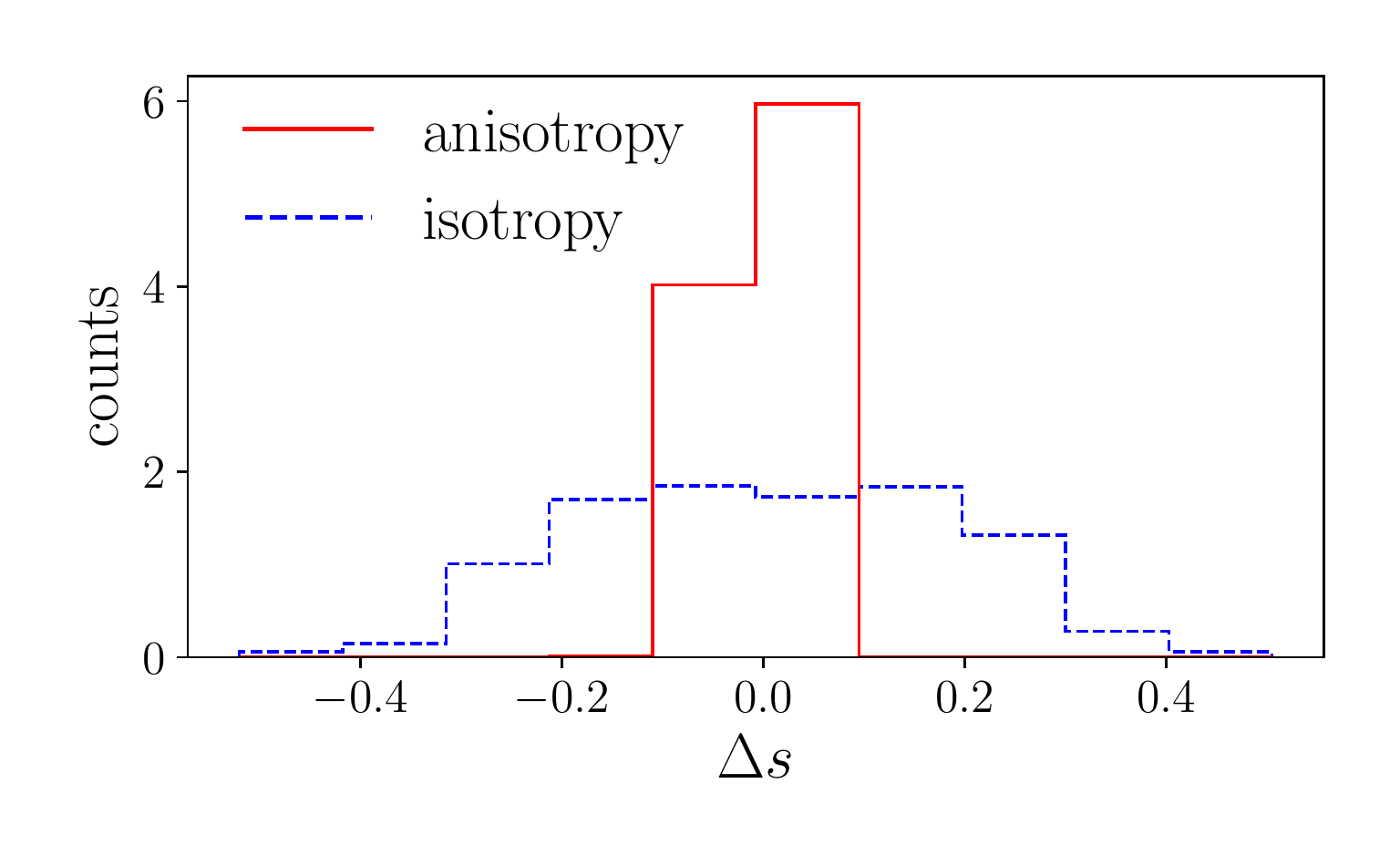}
\subcaption{}
\label{fig:1ddirection_a}
\end{subfigure}
\hspace{25mm}
\begin{subfigure}[b]{0.475\textwidth}
\includegraphics[width=\textwidth]{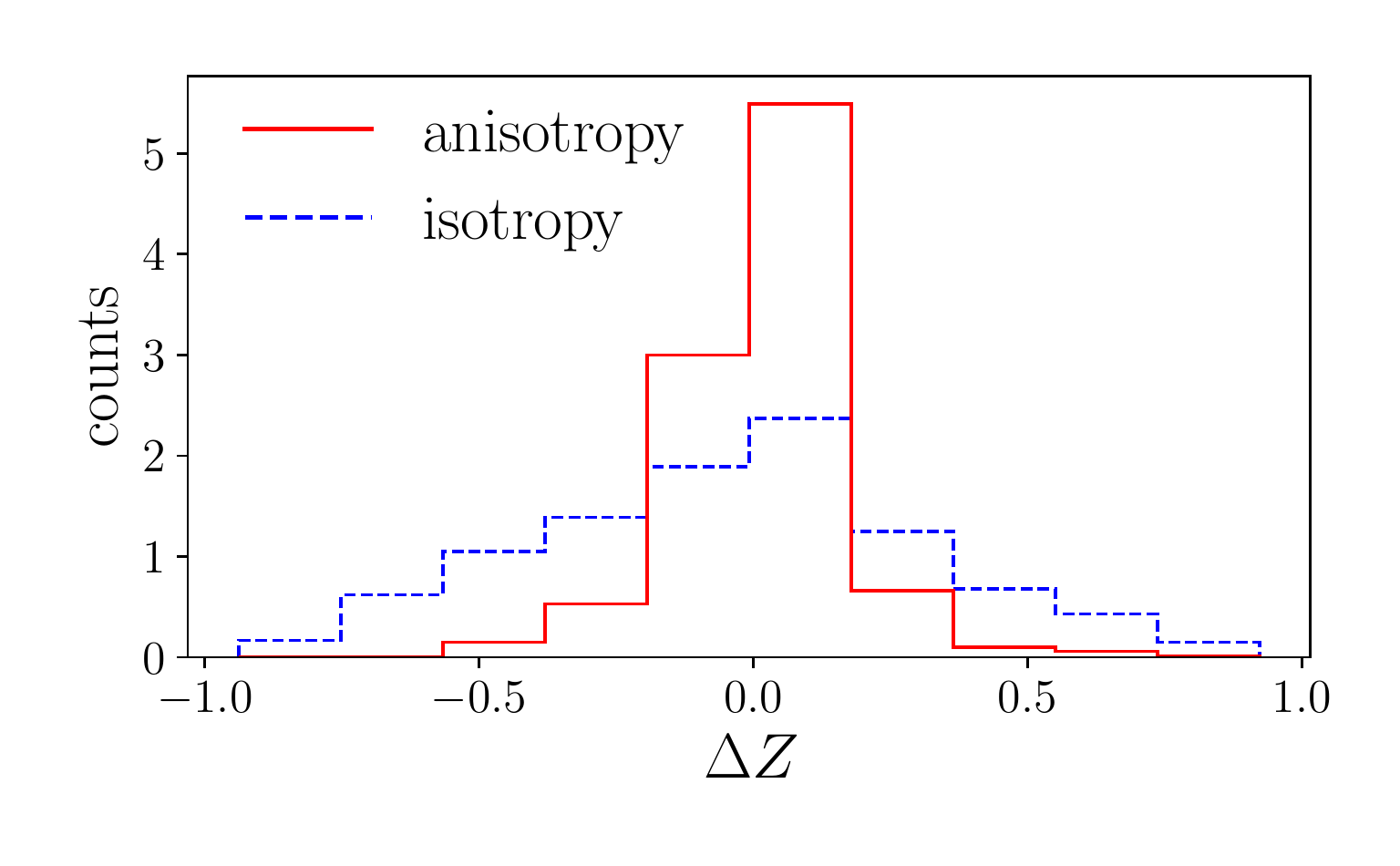}
\subcaption{}
\label{fig:1ddirection_b}
\end{subfigure}
\caption{For each cosmic ray we show a) the directional difference to its source direction, b) the difference between the reconstructed and the true cosmic ray charges. The red solid histograms shows the average distribution of $100$ scenarios with a single signal source and $N=10$ cosmic rays. The blue dashed histograms show the resolution from $100$ isotropic scenarios with $N=1$ cosmic rays from each of the $m=10$ sources.}
\label{fig:1ddirection}
\end{centering}
\end{figure}

We also evaluate the precision in the reconstruction of source directions and cosmic ray charges, again starting with the single-source scenario. In Fig.~\ref{fig:1ddirection_a}, we show the reconstruction quality of the source direction. The red solid histogram represents for each cosmic ray the difference of the reconstructed and true source directions $\Delta s=\hat{s}_i-s_i$. To provide a general measure, we average over $100$ scenarios with $1$ source emitting $N=10$ cosmic rays. The resolution in the source direction is rather good ($\sigma_s=0.024$).

In Fig.~\ref{fig:1ddirection_b}, the red solid histogram also shows the reconstructed cosmic ray charges compared to the true charges $\Delta Z=\hat{Z}_i-Z_i$. Once again, we average over $100$ scenarios. The resolution in the cosmic ray charge amounts to $\sigma_Z=0.15$ for $0\le Z\le 1$ used in the fit. For charges within $c=1\hdots 26$ this implies a resolution of $26\times \sigma_Z\approx 4$ which is superior when compared to the information contained in the $X_{\text{max}}$ measurement which has a logarithmic mass or charge dependence ($Z=A/2$), respectively. Note, however, that the true transformation $T$ in eq.~(\ref{eq:translation}) must be known for this charge reconstruction.

For the isotropic scenario, neither the reconstruction of the source directions nor of the charges work at a satisfactory level. In Fig.~\ref{fig:1ddirection_a}, the blue dashed histogram shows the resolution in the source direction averaged over $100$ isotropic scenarios. In Fig.~\ref{fig:1ddirection_b}, we also show the corresponding charge resolution.
\begin{figure}[ttt]
\captionsetup[subfigure]{aboveskip=-1pt,belowskip=-1pt}
\begin{centering}
\begin{subfigure}[b]{0.38\textwidth}
\includegraphics[width=\textwidth]{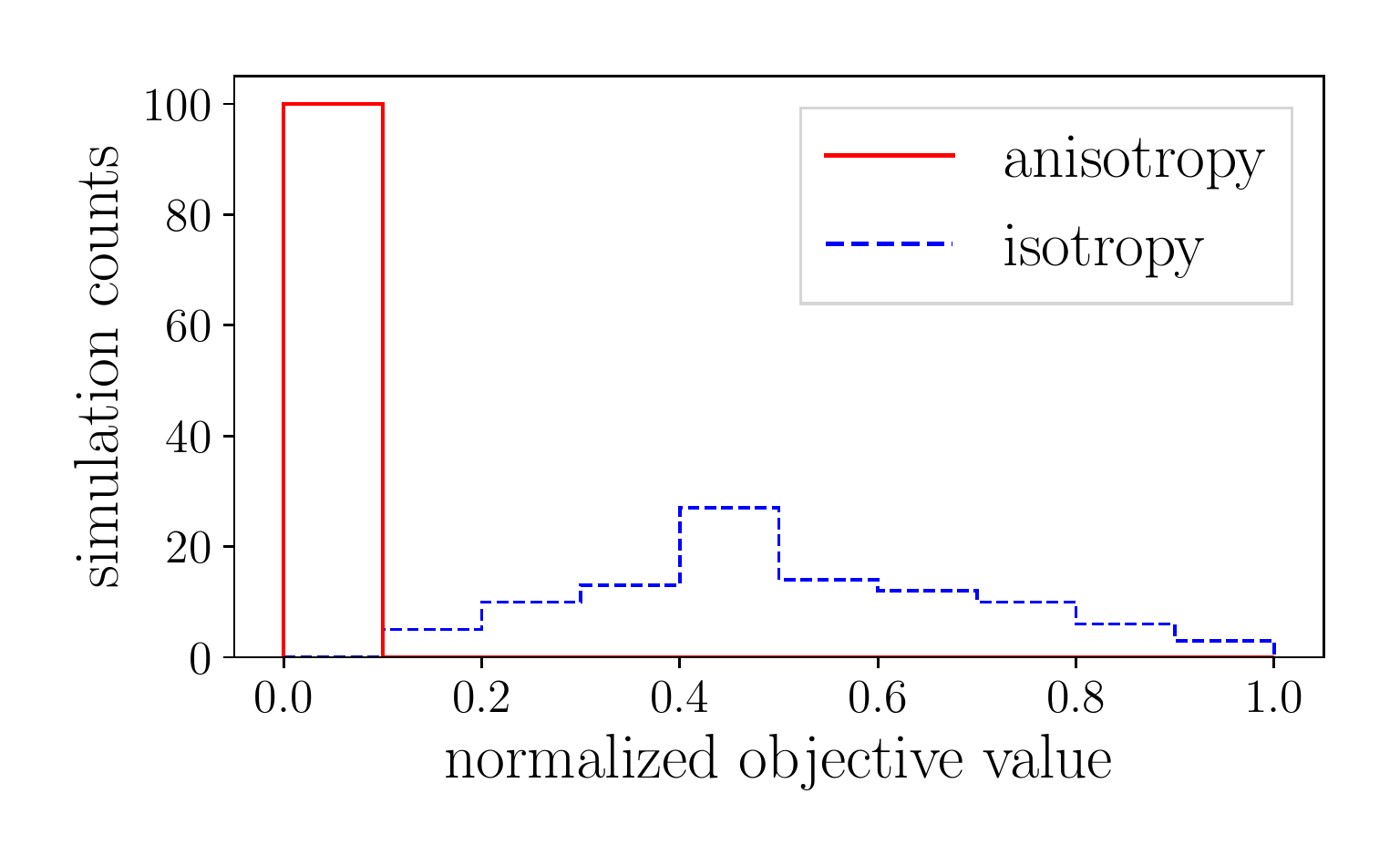}
\subcaption{}
\label{fig:1dsinglesignificance_a}
\end{subfigure}
\hspace{25mm}
\begin{subfigure}[b]{0.38\textwidth}
\includegraphics[width=\textwidth]{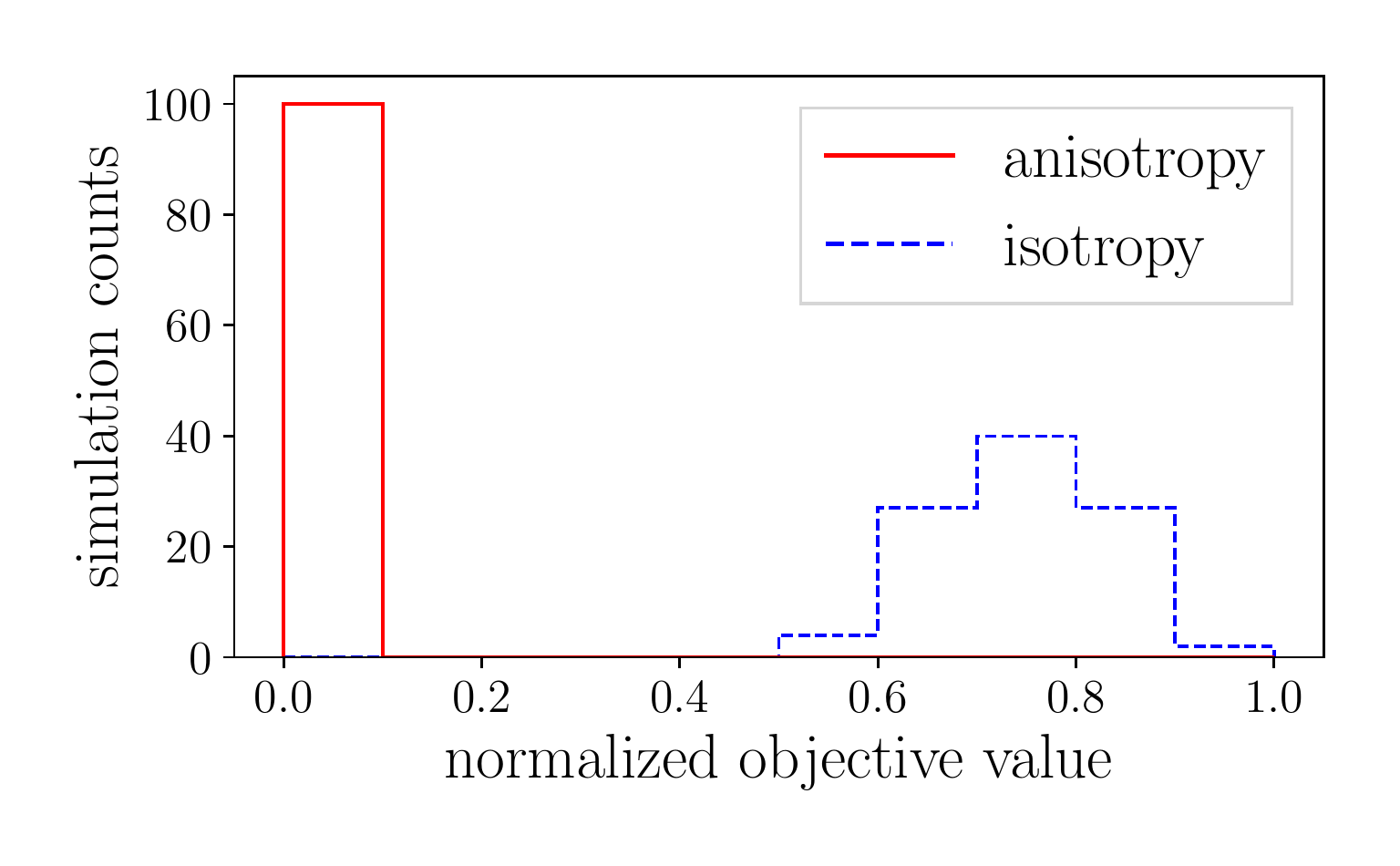}
\subcaption{}
\label{fig:1dsinglesignificance_b}
\end{subfigure}
\hspace{25mm}
\begin{subfigure}[b]{0.38\textwidth}
\includegraphics[width=\textwidth]{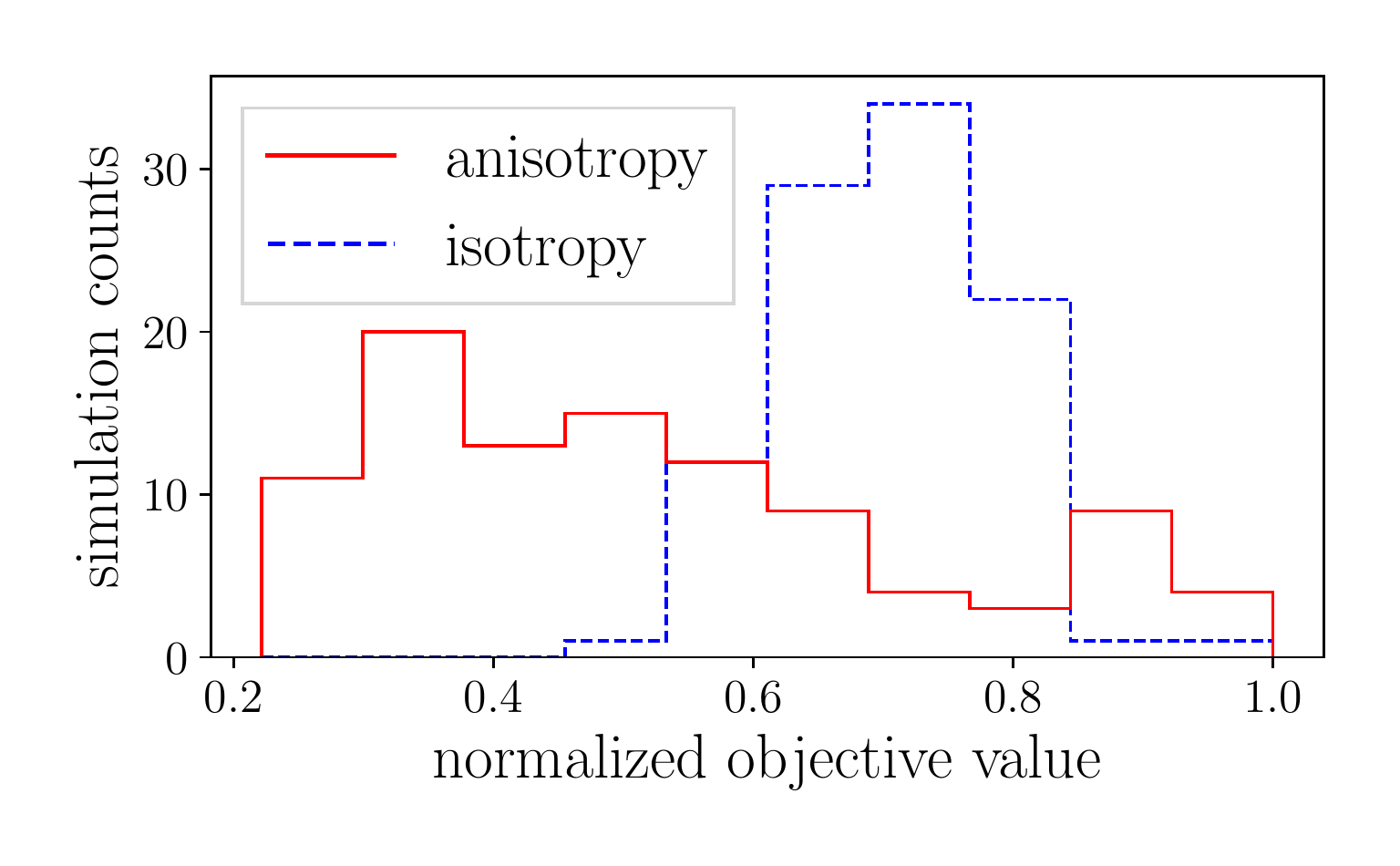}
\subcaption{}
\label{fig:1dsinglesignificance_c}
\end{subfigure}
\caption{Objective values of the fit (\ref{eq:loss}) normalized to the largest observed objective values in the related scenarios. The red solid histograms show the signal scenarios with a single source each, and the blue dashed histograms the corresponding isotropic scenarios, a)~$N=10$ cosmic rays, b)~$N=100$ cosmic rays, c)~$50$ cosmic rays from the source overlayed with $50$ isotropic cosmic rays.}
\label{fig:1dsinglesignificance}
\end{centering}
\end{figure}

In Fig.~\ref{fig:1dsinglesignificance_a}, the red histogram shows the values of the objective function (\ref{eq:loss}) after the fit for $100$ single-source scenarios, and the blue histogram shows the objective values of $100$ isotropic scenarios. For better visibility, we normalized the final objective values to the largest objective value observed in all $200$ scenarios. In this comparison the single-source scenario can be well distinguished from isotropic scenarios.

We also investigate the dependence on the number $N$ of cosmic rays. In Fig.~\ref{fig:1dsinglesignificance_b}, the red solid histogram shows the normalized values of the objective function (\ref{eq:loss}) after the fit to $N=100$ cosmic rays from a single source for $100$ scenarios, and the blue dashed histogram shows the objective values of $100$ isotropic scenarios with $100$ cosmic rays. With increased statistics, the signal scenarios can be even better distinguished from isotropic arrival.

Finally, we study a mixed scenario with a source with $N=50$ cosmic rays which is overlayed with $50$ cosmic rays from an isotropic scenario. As $k$ we used the value $k=N=100$, so the fit tries to find one single source. In Fig.~\ref{fig:1dsinglesignificance_c}, the red solid histogram shows the values of the normalized objective function (\ref{eq:loss}) for the mixed scenario, and the blue dashed histogram shows the objective values of $100$ isotropic scenarios. In spite of the large background, half of the signal scenarios can be distinguished from isotropic arrival.

\begin{figure}[h]
\begin{center}
\includegraphics[width=\linewidth]{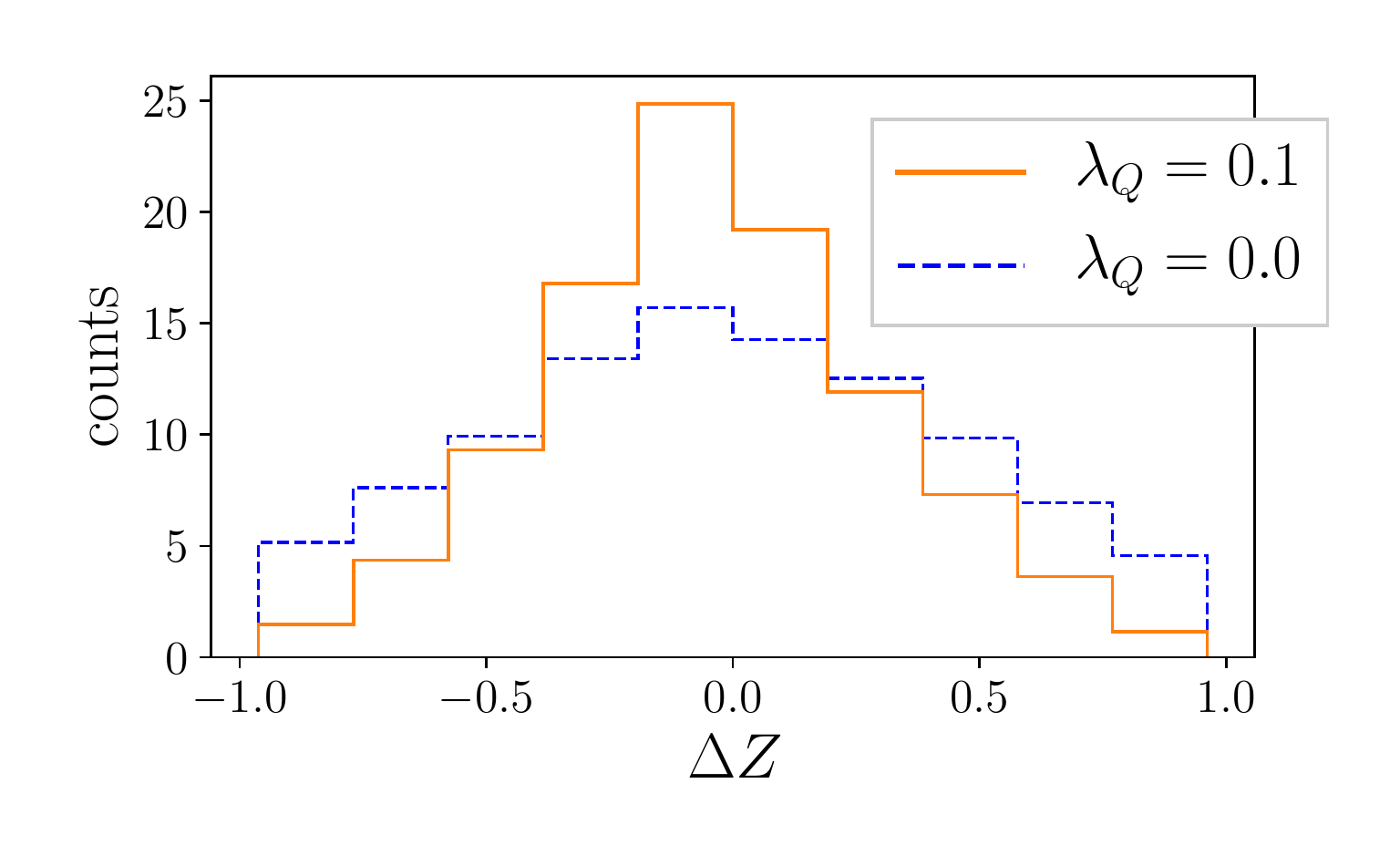}
\caption{Average cosmic ray charge resolution derived from $100$ scenarios with $50\%$ signal and $50\%$ isotropic background with (orange solid histogram, $\lambda_Q=0.1$) and without the charge term contribution to the objective function in (\ref{eq:loss}) (blue dashed histogram).}
\label{fig:1d-Xmax}
\end{center}
\end{figure}

In Fig.~\ref{fig:1d-Xmax} we show the influence of the charge information resulting from measurements of the cosmic ray shower depth $X_{\text{max}}$ for $100$ mixed scenarios with $50\%$ signal and $50\%$ isotropic background. The blue dashed histogram denotes the resolution in the predicted charge $\Delta Z=\hat{Z}_i-Z_i$ obtained when ignoring the charge term (\ref{eq:charge_loss}) in the objective function (\ref{eq:loss}). The orange solid histogram represents the charge resolution when including the charge term in the objective with a hyperparameter $\lambda_Q=0.1$. On average, the resolution improves when including the $X_{\text{max}}$ measurements.

\subsection{Spherical mixed-composition patterns}

In this step, we extend our benchmark studies to the two-dimensional arrival directions of cosmic rays on the surface of a sphere. As our transformation $T$ in eq.~(\ref{eq:fit-parameters}) in the fit, we use the matrix $M(\delta)$ in eq.~(\ref{eq:rotation}) for rotations around the $z$-axis which is perpendicular to the galactic plane. We vary only the longitude $l$ coordinate of the cosmic ray while keeping its latitude $b$ constant. The rotation angle is energy- and charge-dependent according to (\ref{eq:angular_displacement}).

In our simulations, we use uniformly distributed cosmic ray energies between $E_{min} = 40$~EeV and $E_{max} = 100$~EeV. For the charges we assign a uniform distribution between protons ($Z=1$) and iron ($Z=26$). This leads to rotation angles between $\delta=0.02\hdots 1.3$~rad. Again, we use the Gumbel functions $G({A}, E)$ to assign values for the shower depth $X_{\text{max}}$.

\begin{figure}[t]
\begin{center}
\includegraphics[width=\linewidth]{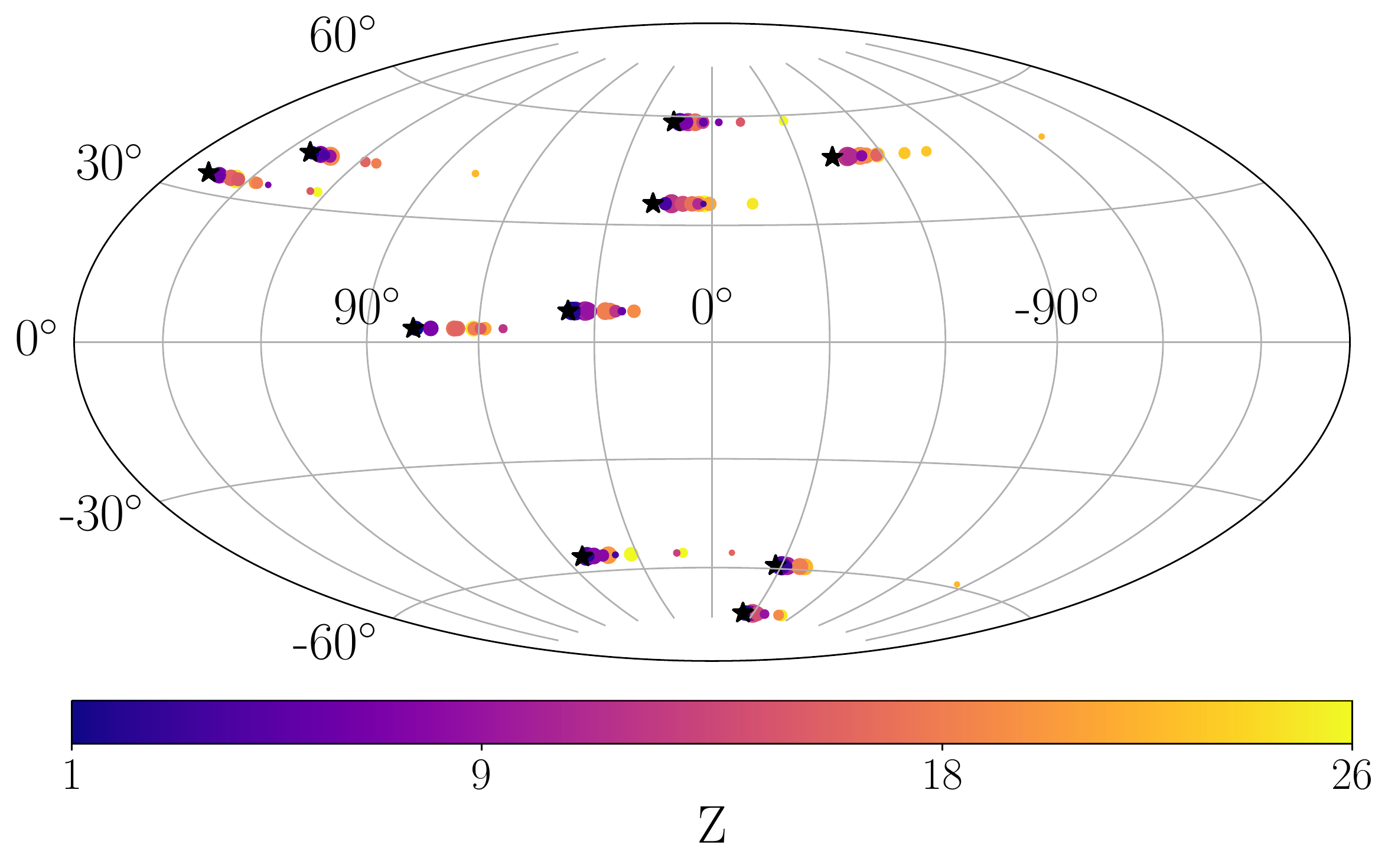}
\caption{Example arrival map of cosmic rays including a coherent rigidity dependent deflection $\delta=-2 Z/(E/\text{EeV})$ as a rotation around the vertical axis. The black stars denote the $m=10$ sources each emitting $N=10$ cosmic rays shown by circular symbols. The symbol size represents the cosmic ray energies between $E=40 \, \text{EeV} \hdots 100 \, \text{EeV}$, and the color scale denotes their charges between $1\le Z\le 26$.} 
\label{fig:sphere}
\end{center}
\end{figure}

Fig.~\ref{fig:sphere} shows an example arrival map for $m=10$ sources each emitting $N=10$ cosmic rays. The symbol sizes indicate the cosmic ray energies, and the charges are denoted by the color code. 

As the initial parameters for the fit in eq.~(\ref{eq:fit-parameters}), for the predicted cosmic ray arrival directions $\hat{s}_i$ outside the galaxy we use the observed arrival directions on Earth $\hat{s}_i=p_i$, and for the charges $\hat{Z}_i$ we apply Bayes' theorem to the shower depth $X_{\text{max}, i}$ assuming a flat composition prior between $\hat{Z}_i = 1\hdots 5$. 

\begin{figure}[ttt]
\captionsetup[subfigure]{aboveskip=-1pt,belowskip=-1pt}
\begin{centering}
\begin{subfigure}[b]{0.475\textwidth}
\includegraphics[width=\textwidth]{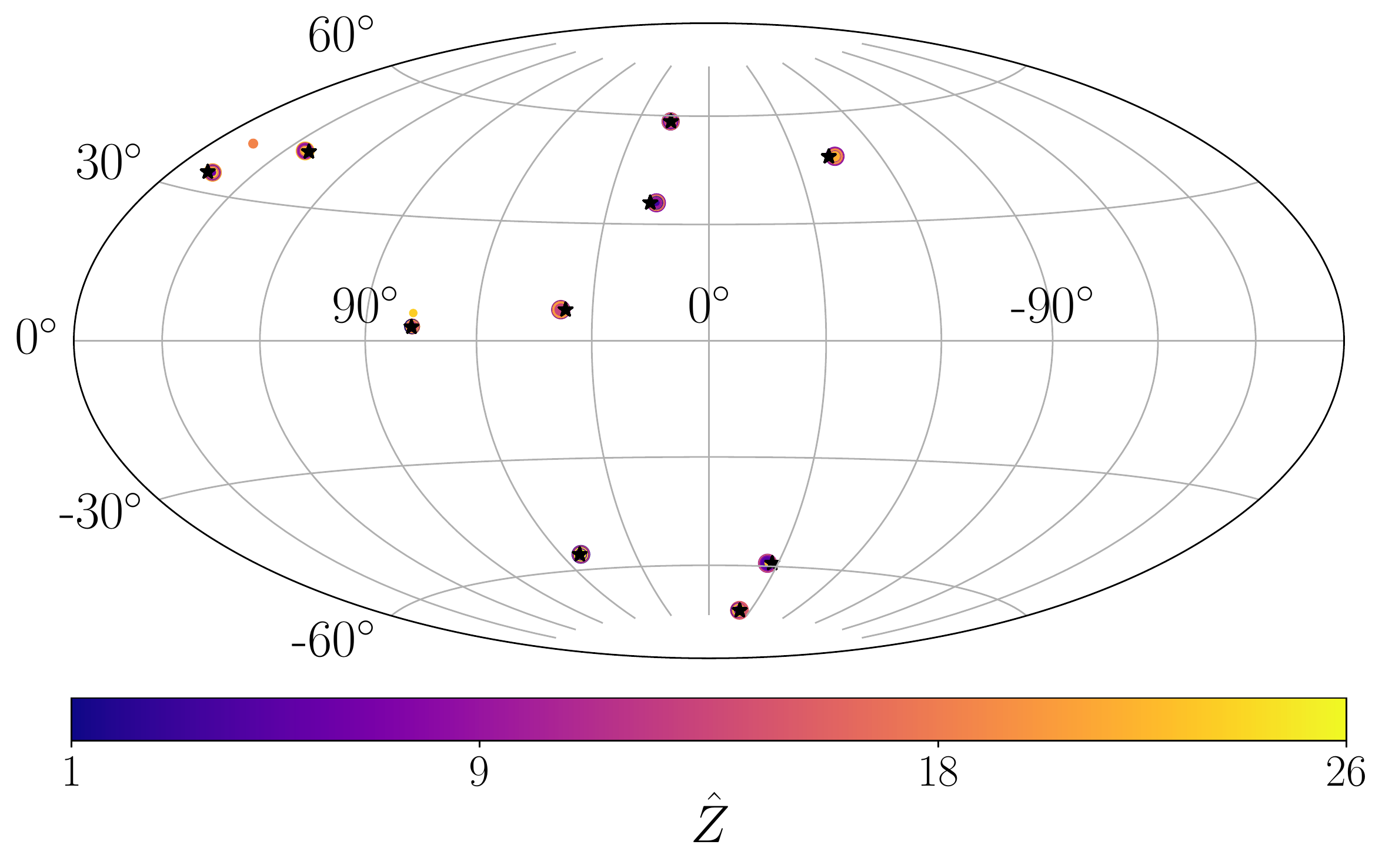}
\subcaption{}
\label{fig:2dmultiplesources_a}
\vspace{3mm}
\end{subfigure}
\hspace{25mm}
\begin{subfigure}[b]{0.475\textwidth}
\includegraphics[width=\textwidth]{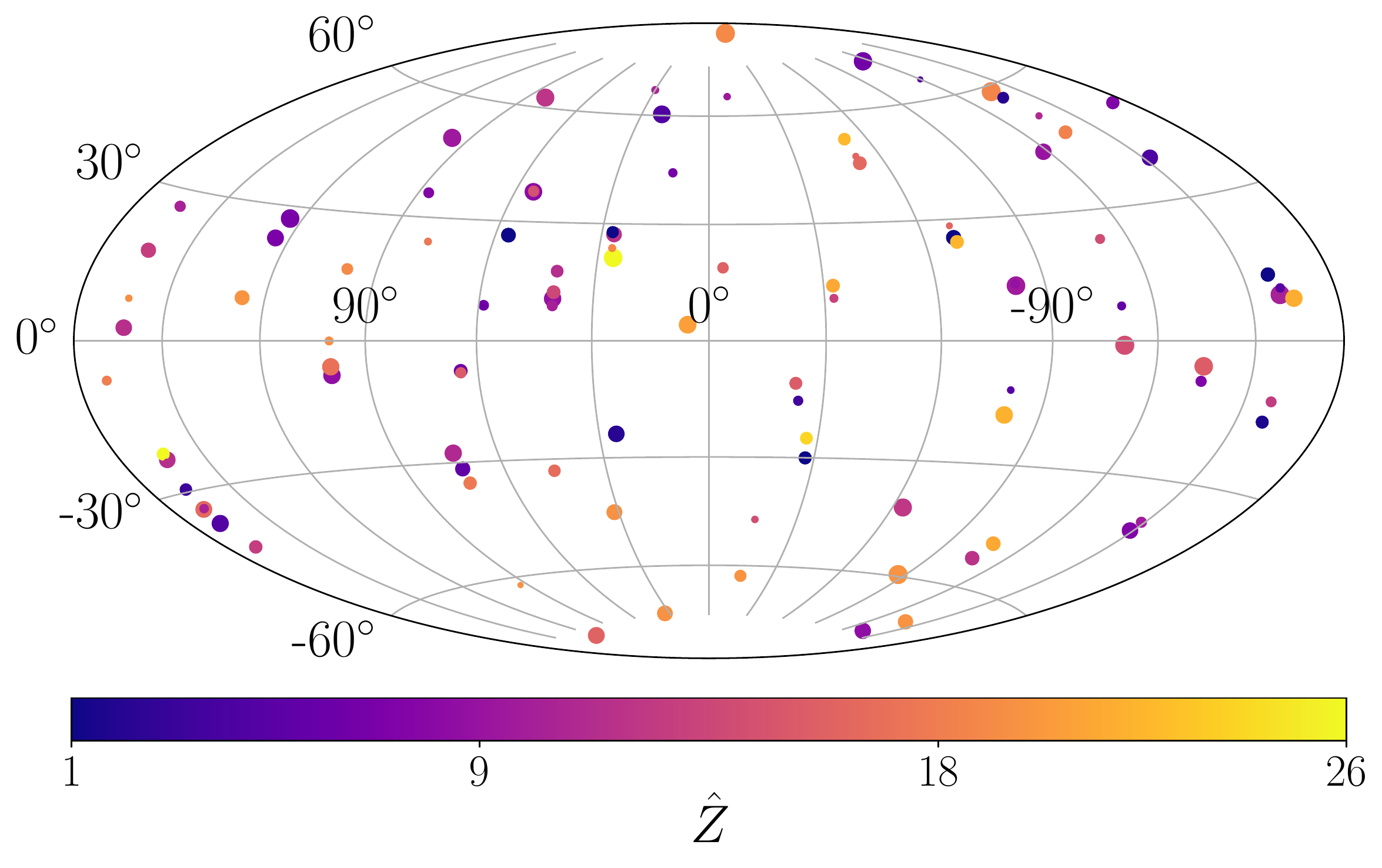}
\subcaption{}
\label{fig:2dmultiplesources_b}
\end{subfigure}
\caption{Fit results: a) Derived original directions of cosmic rays prior to deflection for the signal scenario with $m=10$ sources each emitting $N=10$ cosmic rays shown in Fig.~\ref{fig:sphere}. Here, most of the cosmic ray directions cluster around their source directions. b) Derived original directions of $N=100$ isotropic cosmic rays with many low-occupancy clusters of original directions.}
\label{fig:2dmultiplesources}
\end{centering}
\end{figure}

In Fig.~\ref{fig:2dmultiplesources_a} we show the result of the $2$-dimensional fit for this source scenario. The stars again denote the source directions, and the circular symbols present the predicted arrival directions prior to the coherent deflections. The vast majority of the $N=100$ cosmic rays has been assigned correctly to their source directions and are almost not visible owing to the large clustering strengths at the sources. 

A small fraction of the cosmic rays remains isolated or was allocated to the wrong sources; this is attributable to a combination of the initial charge assignment and the algorithm for evaluating original directions of neighboring cosmic rays. Once the allocation to a group of nearest cosmic rays is incorrect, it appears to be difficult for the fit to reassign the cosmic ray to another group of neighbors.

\begin{figure}[ttt]
\captionsetup[subfigure]{aboveskip=-1pt,belowskip=-1pt}
\begin{centering}
\begin{subfigure}[b]{0.45\textwidth}
\includegraphics[width=\textwidth]{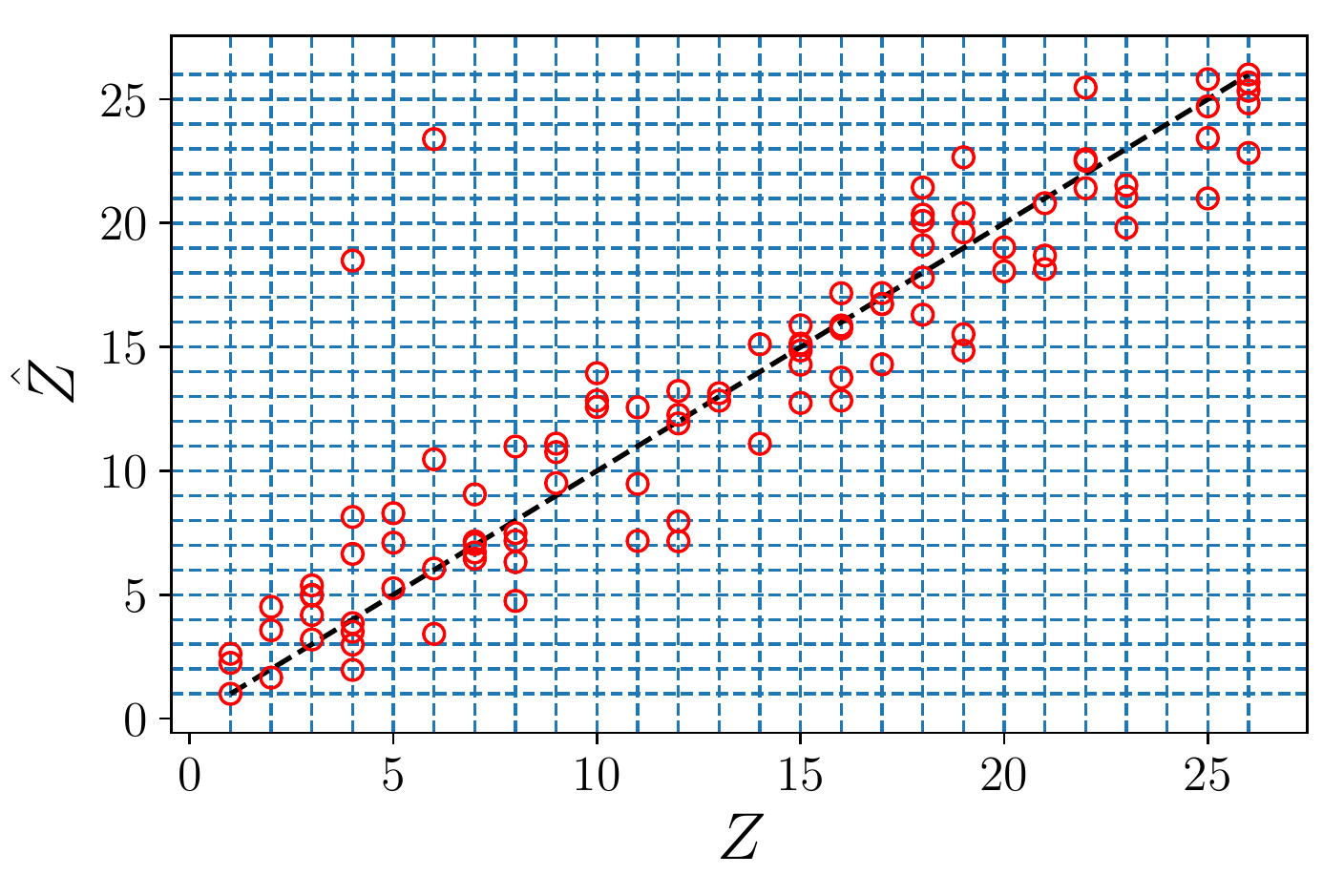}
\subcaption{}
\label{fig:2dmultiplecharges_a}
\end{subfigure}
\hspace{25mm}
\begin{subfigure}[b]{0.45\textwidth}
\includegraphics[width=\textwidth]{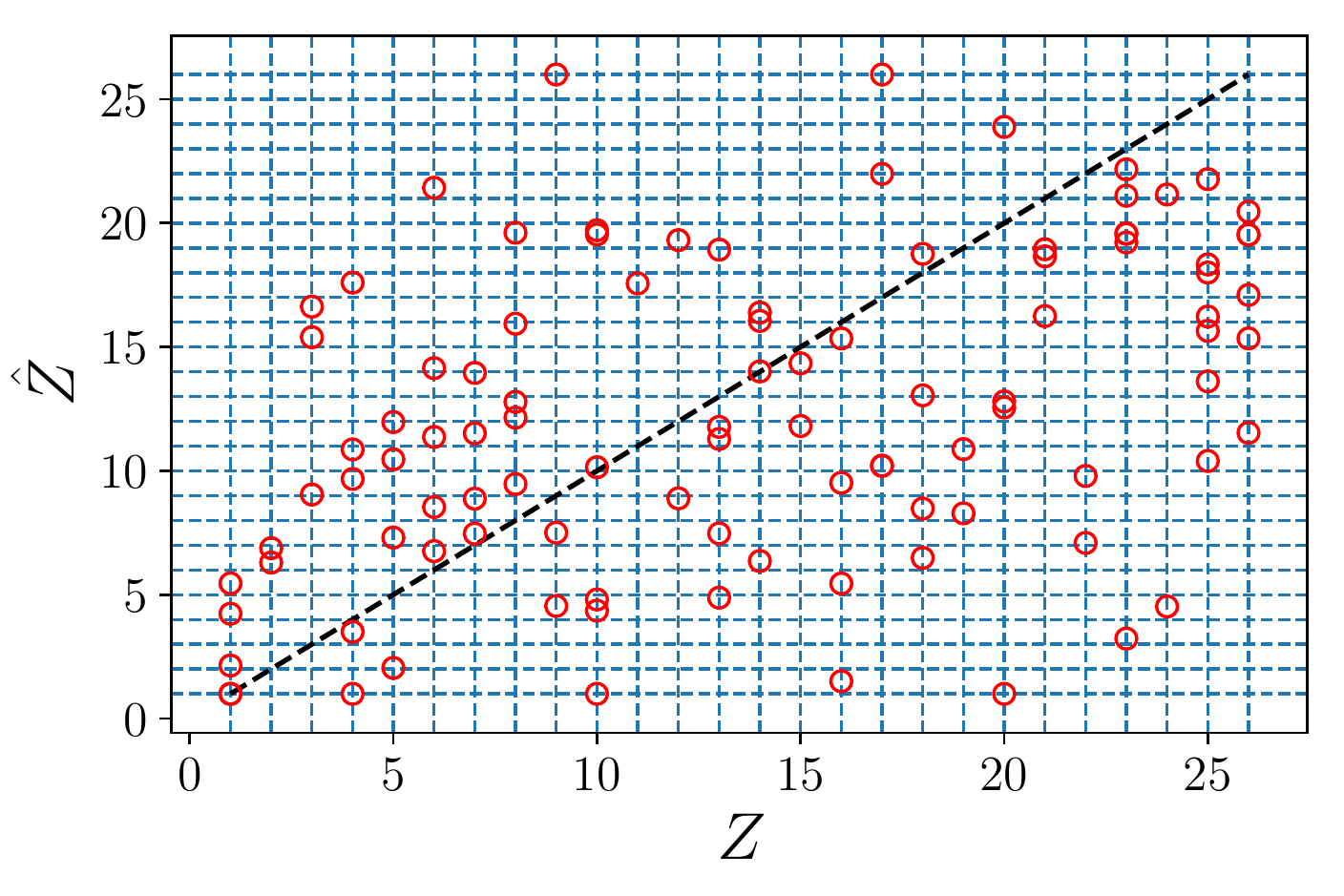}
\subcaption{}
\label{fig:2dmultiplecharges_b}
\end{subfigure}
\caption{Reconstructed cosmic ray charges compared to the true charges a) for the signal scenario with $m=10$ sources each emitting $N=10$ cosmic rays shown in Figs.~\ref{fig:sphere} and \ref{fig:2dmultiplesources_a}, b) for the isotropic scenario with $N=100$ cosmic rays shown in Fig.~\ref{fig:2dmultiplesources_b}.}
\label{fig:2dmultiplecharges}
\end{centering}
\end{figure}

In Fig.~\ref{fig:2dmultiplecharges_a} we show the reconstructed charges of the cosmic rays as a function of their true charges. A linear correlation of the reconstructed with the true charges is visible for the majority of the cosmic rays. The two outliers can be identified in Fig.~\ref{fig:2dmultiplesources_a}, one in the upper left in an equilibrium between two source positions, and one at longitude $80$~deg and latitude $5$~deg misidentified with a different source due to the high assigned charge.

\begin{figure}[h]
\begin{center}
\includegraphics[width=0.9\linewidth]{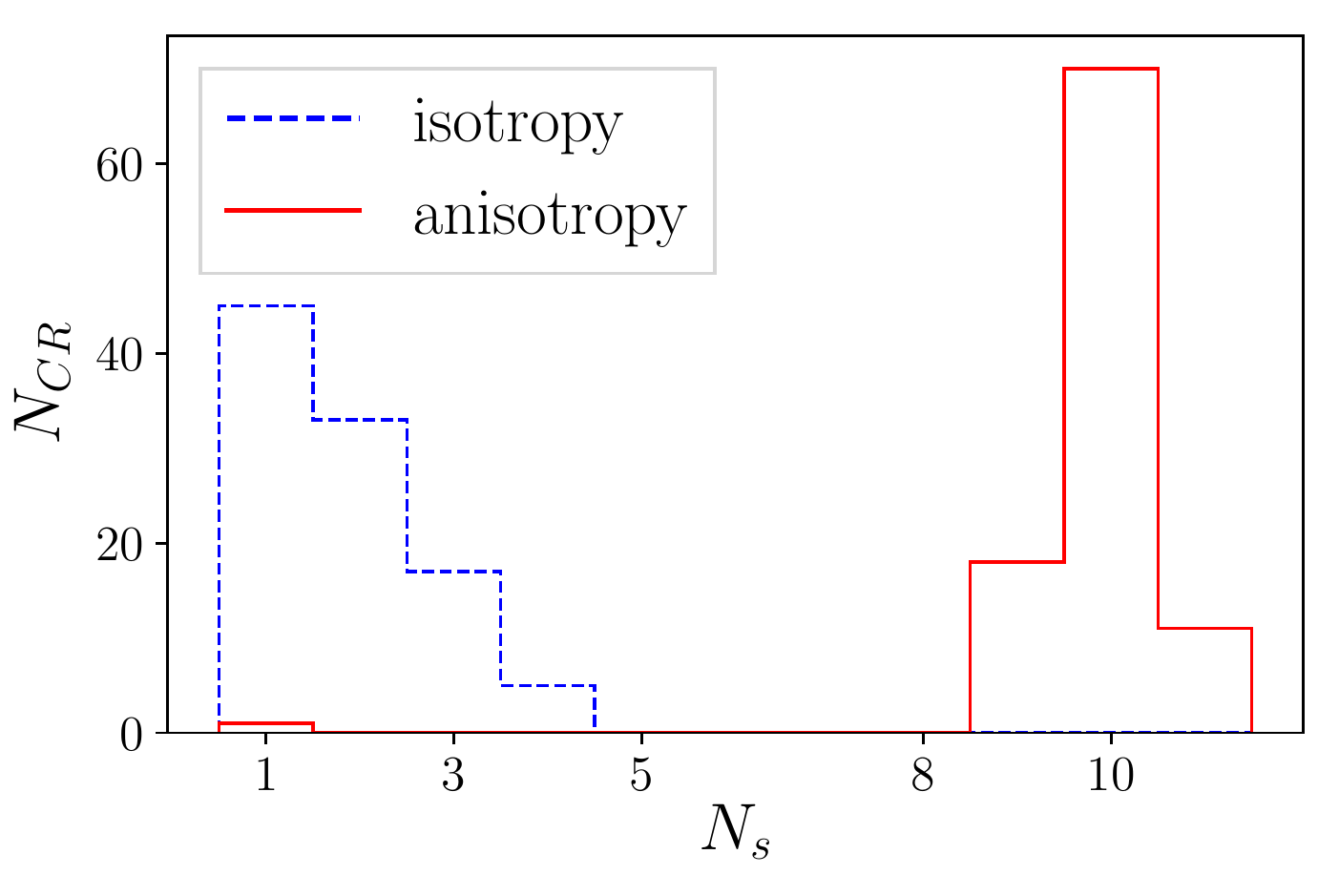}
\caption{Cluster strength measured in terms of top-hat counting: number of cosmic ray original directions $\hat{s}$ predicted by the fit within a radial distance of $5$ deg from the original direction $\hat{s}_i$ of each cosmic ray. The red solid histogram represents the signal scenario with $m=10$ sources each emitting $N=10$ cosmic rays shown in Figs.~\ref{fig:sphere} and \ref{fig:2dmultiplesources_a}, the blue dashed histogram denotes the isotropic scenario with $N=100$ cosmic rays shown in Fig.~\ref{fig:2dmultiplesources_b}. }
\label{fig:2dtophat}
\end{center}
\end{figure}

As a measure of clustering strength we use the so-called top-hat counting \cite{alexandreas}. For each cosmic ray we count the number of cosmic rays within a radial distance of $5$ deg and include the initiating cosmic ray in the count. In Fig.~\ref{fig:2dtophat} the red solid histogram shows the count for each of the $N=100$ cosmic rays. The majority of the cosmic rays yield a cluster strength of $10$, as expected, when they are correctly allocated to their sources. The above-mentioned cosmic ray which was incorrectly allocated to another source is visible at $N_s=11$, and also the cosmic ray that remained in an equilibrium between two sources is visible at $N_s=1$.

For comparison, we also present the results of a fit to an isotropic scenario (the original map is not shown here). In Fig.~\ref{fig:2dmultiplesources_b} we show the predicted original directions of the cosmic rays prior to their deflections. The fit attempts to concentrate the predicted directions as required in the objective function (\ref{eq:loss}), however, with the result of many small clusters spread around the sphere. 

In Fig.~\ref{fig:2dtophat} the blue dashed histogram shows the clustering strength by the top-hat counting explained above. Compared to the $m=10$ source scenario, only small clusters are formed. Therefore, the source scenario can be easily separated from isotropic scenarios, and top-hat counting can serve to evaluate the significance of a signal scenario.

In Fig.~\ref{fig:2dmultiplecharges_b}, we also show the reconstructed charges of the cosmic rays for the isotropic scenario. Here, a weak correlation between the true charges $Z$ and the reconstructed charges $\hat{Z}$ is found which results from the $X_{\text{max}}$ constraint of the charge objective (\ref{eq:charge_loss}). 

\begin{figure*}[ttt]
\begin{center}
\includegraphics[width=0.8\textwidth]{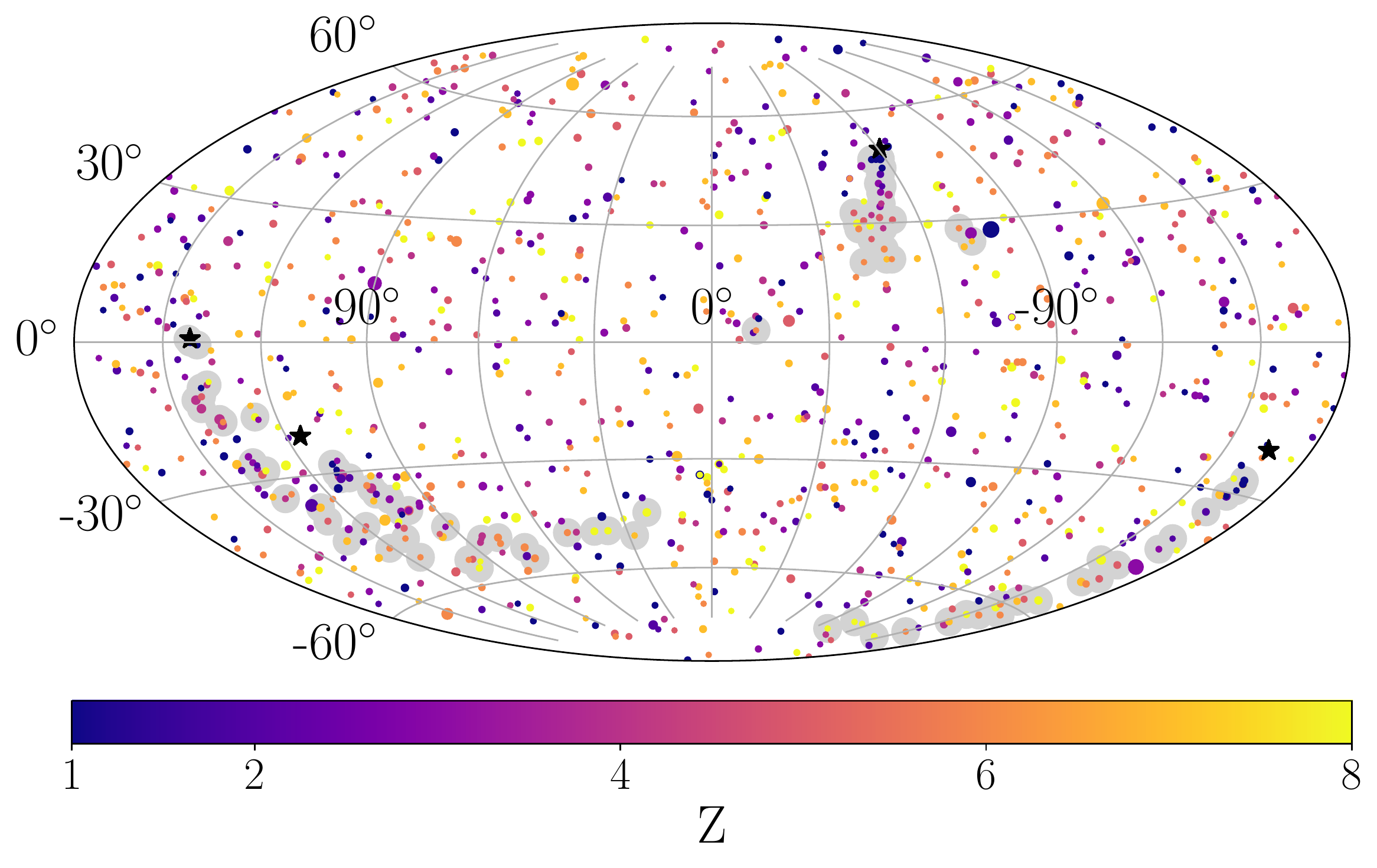}
\caption{Arrival map of $N=1000$ cosmic rays from an astrophysical scenario with $4$ sources (star symbols) each emitting $25$ cosmic rays, and $900$ additional isotropic cosmic rays. The mixed composition consists of nuclei reaching from protons to oxygen. Deflections are performed by the JF12 parameterization of the regular galactic magnetic field with an additional Gaussian spread of size $\sigma = 0.5 \cdot Z / (E/\text{EeV}) \, \text{rad}$. The symbols represent cosmic rays with energies $E\ge 40$~EeV. The symbol size corresponds to the cosmic ray energy, and the color scale denotes their charges. The gray shaded areas indicate cosmic rays originating from the sources.}
\label{fig:GMF}
\end{center}
\end{figure*}

\section{Cosmic ray sources by correcting for galactic magnetic field deflections 
\label{sec:GMF-sources}}

Finally, we apply the fit to correct for cosmic ray deflections in the galactic magnetic field. As our simulated astrophysical scenario, we use $m=4$ sources each emitting $N=25$ cosmic rays and an additional $900$ cosmic rays following an isotropic distribution. The energy spectrum follows \cite{Abraham:2010mj} above a lower-energy cut-off of $E>40$~EeV. With a total of $1000$ cosmic rays above this energy threshold, we yield cosmic ray statistics compatible with actual experiments.

We use an energy-independent mixed composition between proton and oxygen ($Z=1, \hdots, 8$) that follows a uniform charge distribution. Also here, we use the Gumbel functions $G({A}, E)$ to assign shower depths $X_{\text{max}}$ to each cosmic ray. The upper charge limit is meant to avoid low cosmic ray rigidities $R=E/Z$ which, depending on the arrival direction, may lead to large non-ballistic deflections \cite{Erdmann:2016vle}. 

Cosmic ray deflections in the galactic magnetic field are performed using the regular JF12 field \cite{Jansson2012a} encoded in rigidity-dependent magnetic lenses \cite{Bretz2014, Batista2016}. The lenses are based on HEALPix coordinates using $1$~deg$^2$ resolution \cite{Gorski2005}. In order to include a turbulent magnetic field component, we add a Gaussian smearing of the arrival directions with a standard deviation of
\begin{equation}
\sigma = 0.5 \, \frac{Z}{E/\text{EeV}} \, \text{rad.}
\label{eq:blurring}
\end{equation}

In Fig.~\ref{fig:GMF} we show the arrival directions of the cosmic rays on Earth where the circular symbols denote the cosmic rays, the sizes represent the cosmic ray energies, and the color codes denote the charges. The gray shaded areas indicate cosmic rays originating from the sources which are marked by star symbols.

In our fit we again use the regular field of the JF12 parameterization \cite{Jansson2012a} to calculate cosmic ray deflections. To address cosmic ray deflections in the fit, we replaced the lenses by a deep neural network $L$ which serves to predict magnetic field deflections in eq.~(\ref{eq:GMF}). For a given cosmic ray direction $\hat{s}_i$ outside the galaxy, energy $E_i$ and charge $\hat{Z}_i$, the network outputs the most likely arrival direction on Earth, thus providing the transformation $T=L$ in eq.~(\ref{eq:fit-parameters}). We verified that the network provides a good representation of the lenses above cosmic ray rigidities $R=1$~EV, and interpolates well with respect to continuous rigidity and arrival direction values. 

\begin{figure*}[tt]
\begin{center}
\includegraphics[width=0.8\textwidth]{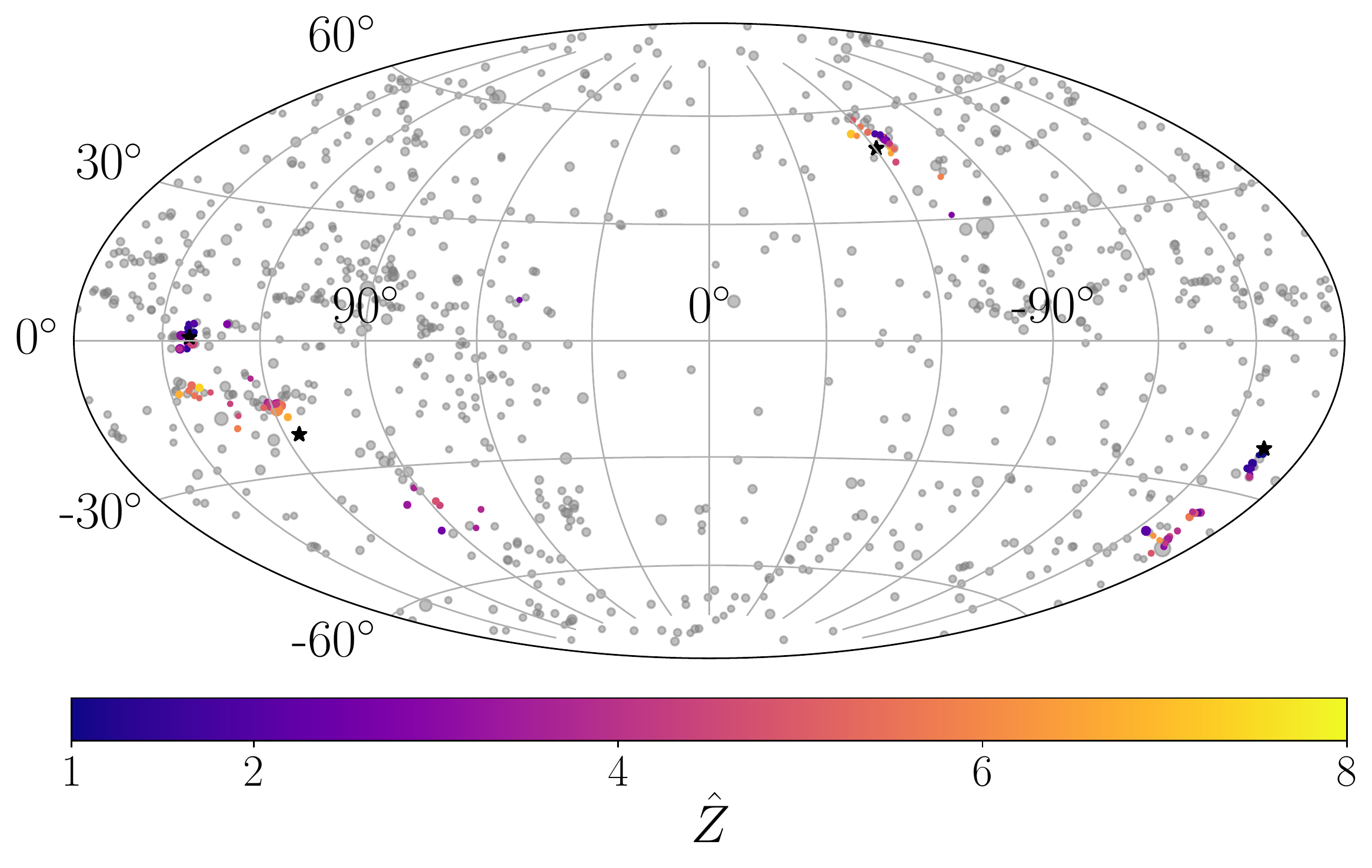}
\caption{Fit results of the derived original directions $\hat{s}$ of cosmic rays prior to deflection in the galactic magnetic field for the source scenario shown in Fig.~\ref{fig:GMF}. The colored symbols denote cosmic rays originating from the sources, and the color code represents their corresponding fitted charge. The gray symbols show the isotropic background contribution.
}
\label{fig:GMF-fit}
\end{center}
\end{figure*}

To take into account turbulent field components or possible uncertainties in the regular field, we adapted the transverse size of the elliptical search region to $4$~deg; details can be found around eq.~(\ref{eq:center-pot-neighbor}) in section \ref{sec:objective} above.

For the fit, initial values of the predicted cosmic ray arrival directions $\hat{s}_i$ outside the galaxy and the charges $\hat{Z}_i$ in eq.~(\ref{eq:fit-parameters}) are required.
We set $\hat{s}_i$ to the observed arrival directions $\hat{s}_i=p_i$. To estimate the cosmic ray charges we apply Bayes' theorem to the shower depth $X_{\text{max}, i}$ assuming a flat composition prior between $\hat{Z}_i = 1\hdots 5$. 

Fig.~\ref{fig:GMF-fit} shows the fit results of the original arrival directions $\hat{s}_i$ for all cosmic rays. Most of the signal cosmic rays (colored symbols) cluster in regions close to their original sources. 

Isotropic background events (gray shading) tend to cluster as well, as required by the fit (\ref{eq:center-pot-neighbor}), where some directions of the sky are more populated than others. The latter effect results from the directional dependent transparency of the galactic magnetic field model. Some of the signal events are attracted by clusters of background events and therefore miss their original source directions. Conversely, the source regions contain some background cosmic rays that are pulled towards the high clustering region.

To quantify a significance for fit results of source scenarios with $m=4$ sources, we investigate the distribution of top-hat counts within $5 \deg$. To account for the above-mentioned directional dependent transparency of the galactic magnetic field model, we show the top-hat counts $N_s$ relative to the average top-hat counts expected for isotropic distributions $\langle N_s^{\text{iso}} \rangle$ at the same arrival direction, i.e., the same HEALPix coordinate at $1$~deg$^2$ resolution.

\begin{figure}[ttt]
\captionsetup[subfigure]{aboveskip=-1pt,belowskip=-1pt}
\begin{centering}
\begin{subfigure}[b]{0.43\textwidth}
\includegraphics[width=\textwidth]{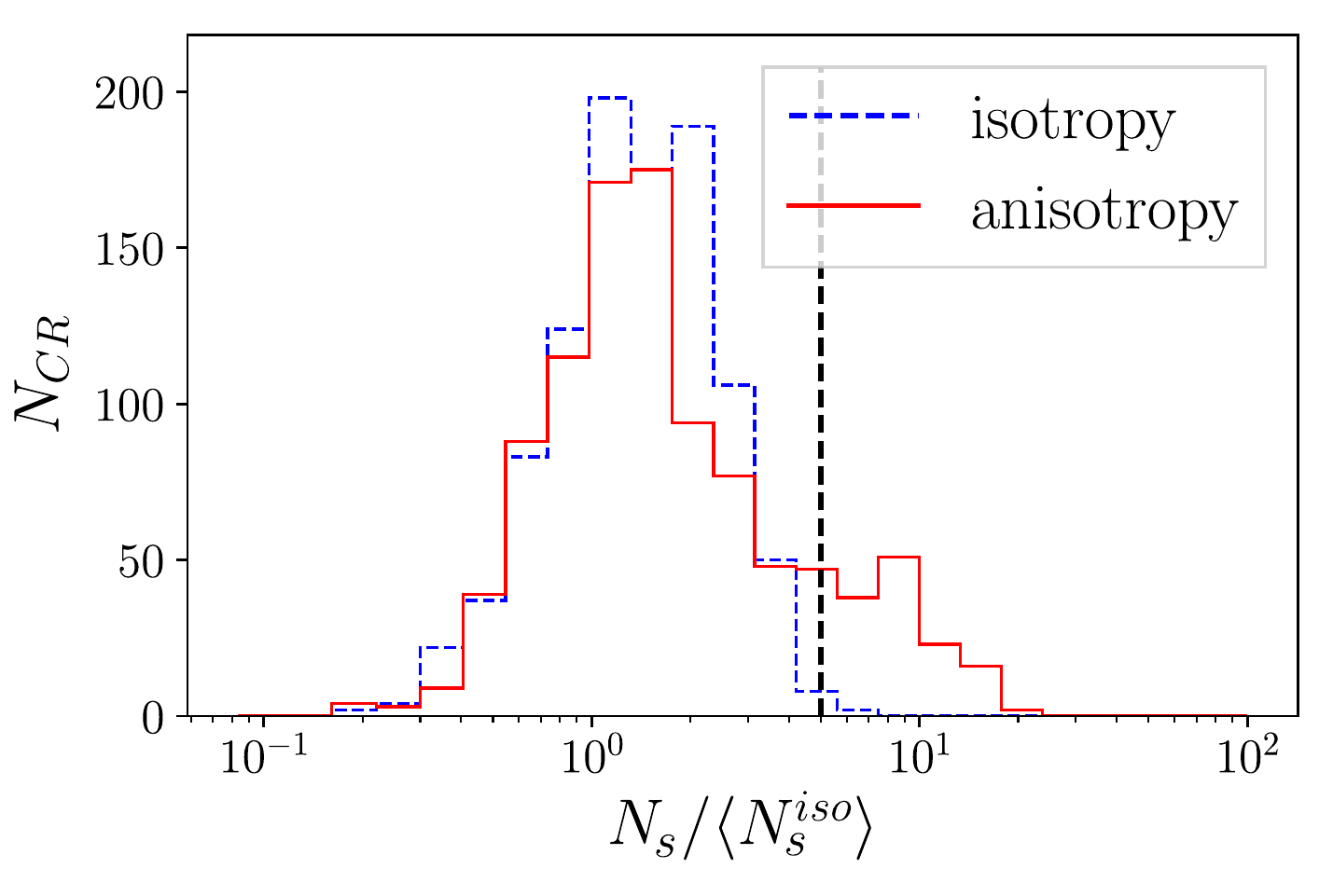}
\subcaption{}
\label{fig:GMFsignificance_a}
\end{subfigure}
\hspace{25mm}
\begin{subfigure}[b]{0.43\textwidth}
\includegraphics[width=\textwidth]{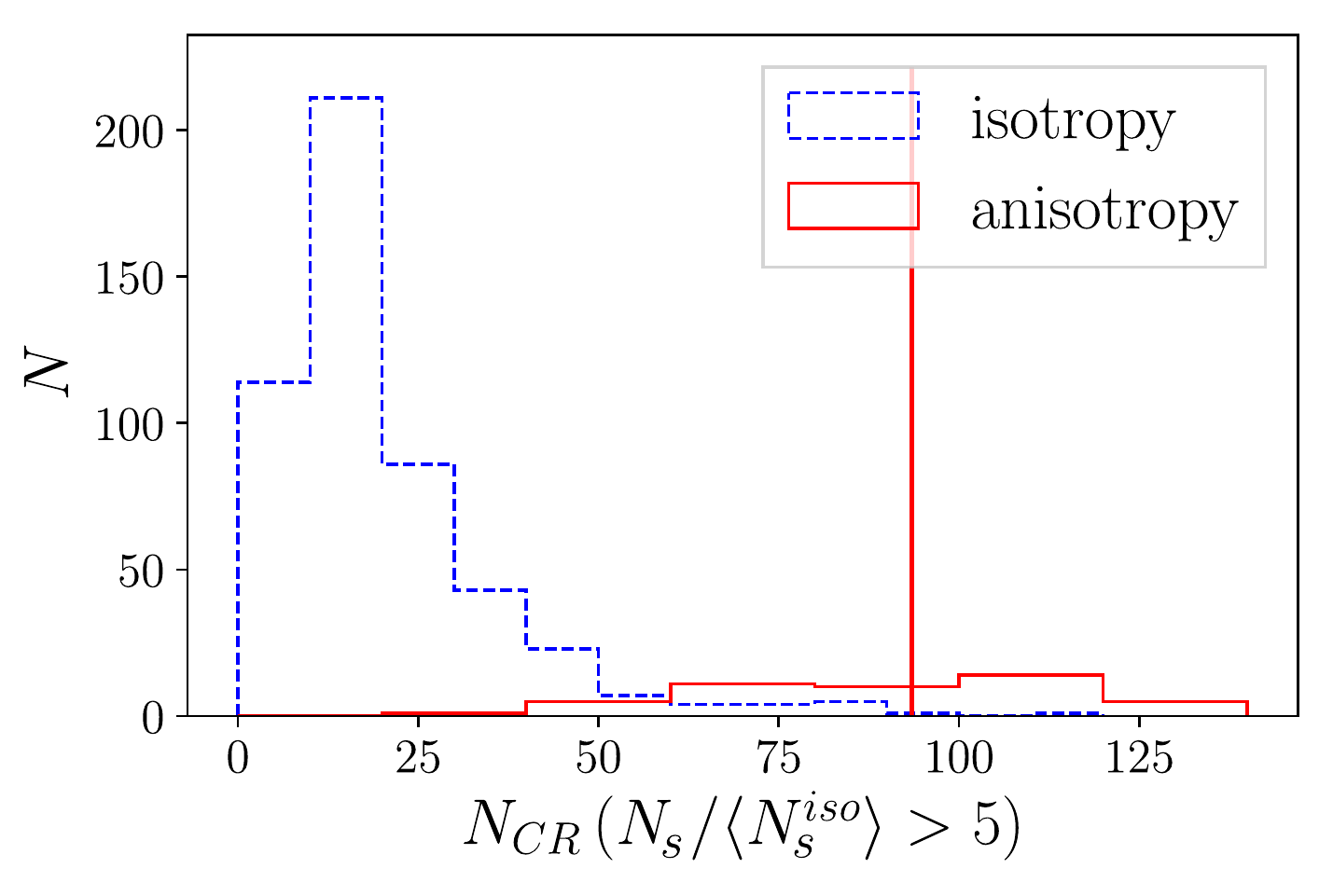}
\subcaption{}
\label{fig:GMFsignificance_b}
\end{subfigure}
\caption{a) Top-hat counts $N_s$ after the fit of the source scenario of Fig.~\ref{fig:GMF-fit}, normalized to the 
average top-hat counts $\langle N_s^{\text{iso}} \rangle$ of isotropic scenarios at the same arrival directions (red solid histogram). The dashed blue histogram shows a fit to an isotropic scenario, and the black vertical line marks $N_s / \langle N_s^{\text{iso}} \rangle = 5$. b)~Number of cosmic rays that exceed the isotropic expectation of top-hat counts by more than a factor of $5$. The red solid histogram shows $50$ astrophysical signal scenarios, containing $m=4$ sources each emitting $25$ cosmic rays overlayed with $900$ isotropic cosmic rays, and the red vertical line denotes the median of the distribution. The blue dashed histogram shows the resulting distribution for $500$ realizations each containing $1000$ cosmic rays with isotropic arrival directions.}
\label{fig:GMFsignificance}
\end{centering}
\end{figure}

In Fig.~\ref{fig:GMFsignificance_a}, we show for the fitted source scenario presented in Fig.~\ref{fig:GMF-fit} the top-hat ratio $N_s/\langle N_s^{\text{iso}} \rangle$ for each cosmic ray using the red solid histogram. To enable a comparison with purely isotropic arrival directions, we also show the fit result of a single isotropic realization using the blue dashed histogram. For the source scenario, a population of cosmic rays is visible exceeding $N_s / \langle N_s^{\text{iso}} \rangle = 5$ (black dashed line). This population corresponds mostly to signal cosmic rays and a few background events which were attracted to the same clusters.

As a statistical measure to discriminate between isotropic and anisotropic scenarios, we use the number of cosmic rays that exceed the threshold $N_{CR} (N_s / \langle N_s^{\text{iso}} \rangle > 5)$ after the fit. In Fig.~\ref{fig:GMFsignificance_b} we show the number of cosmic rays that exceed this threshold for $50$ signal scenarios using the red solid histogram, and its median by the red vertical line. These signal scenarios again contain $m=4$ random source positions and $25$ signal cosmic rays per source, overlayed with $900$ cosmic rays which follow an isotropic distribution. 

For comparison, the blue dashed histogram represents the expectations for $500$ isotropic realizations. We find that a single isotropic set exceeds the median expectation for our astrophysical scenario. Thus, as the expected sensitivity to distinguish clustering from an astrophysical scenario with $10\%$ signal fraction from $4$ sources from clustering resulting from isotropic arrival distributions we report a p-value of $p = 2 \cdot 10^{-3}$ corresponding to $2.9$ standard deviations.

\begin{figure}[ttt]
\captionsetup[subfigure]{aboveskip=-1pt,belowskip=-1pt}
\begin{centering}
\begin{subfigure}[b]{0.48\textwidth}
\includegraphics[width=\textwidth]{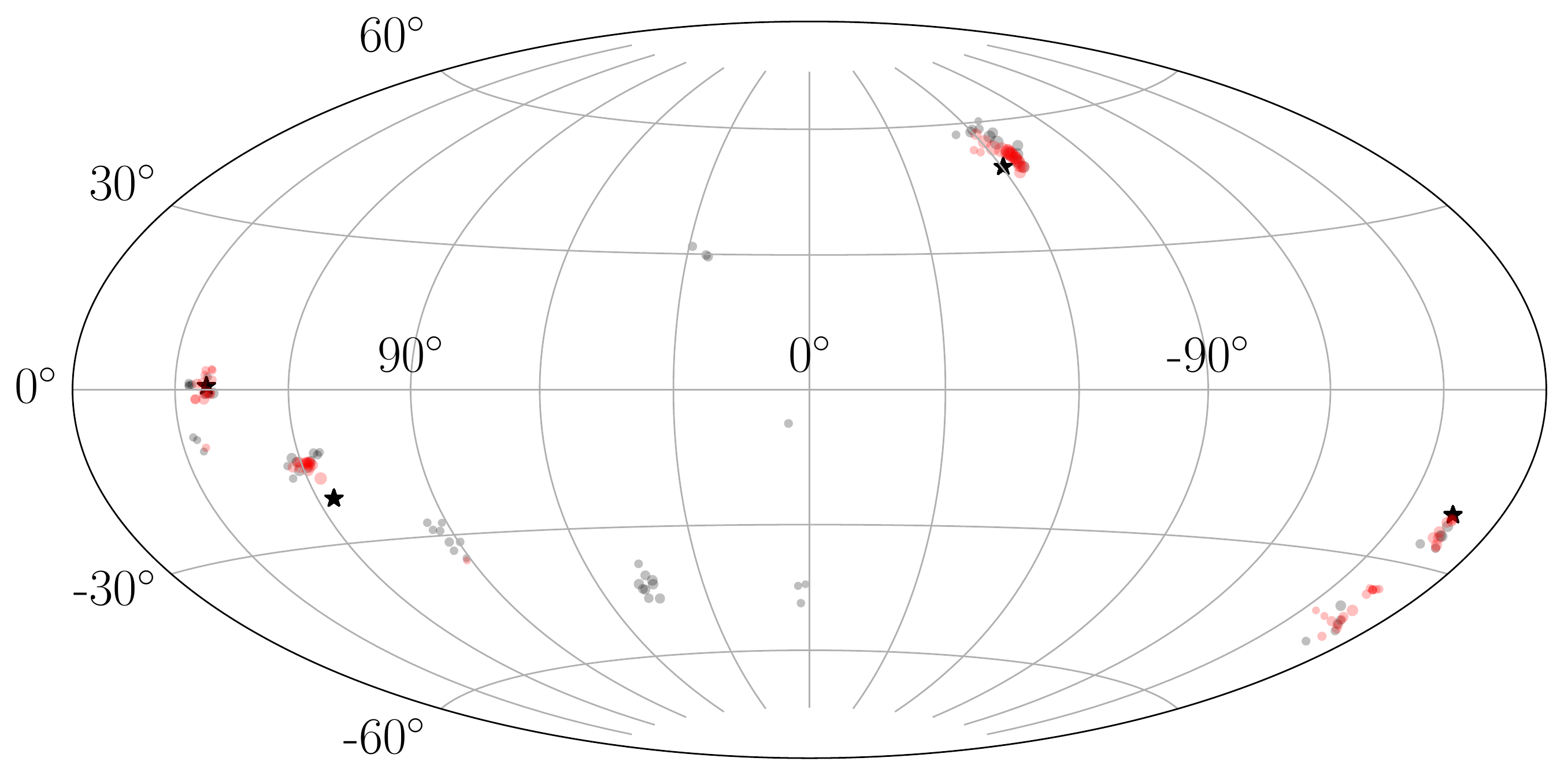}
\subcaption{}
\label{fig:GMF-signal_a}
\vspace{8mm}
\end{subfigure}
\hspace{25mm}
\begin{subfigure}[b]{0.43\textwidth}
\includegraphics[width=\textwidth]{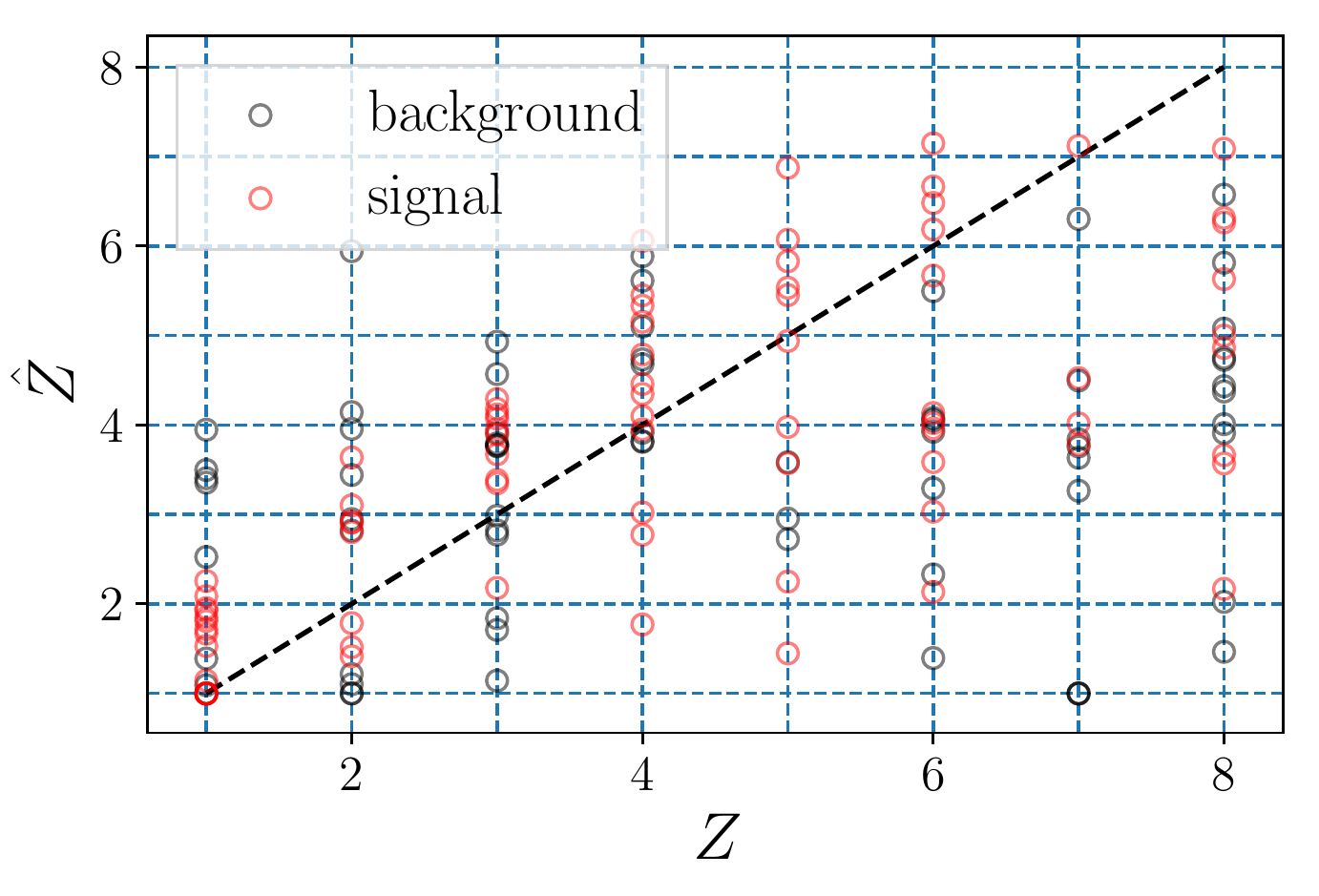}
\subcaption{}
\label{fig:GMF-signal_b}
\end{subfigure}
\caption{Reconstruction quality for signal enhanced cosmic rays ($N_s / \langle N_s^{\text{iso}} \rangle > 5$) from the fit result shown in Fig.~\ref{fig:GMF-fit}. Red symbols represent the signal cosmic rays and gray symbols the isotropic background contribution. a) Fitted cosmic ray source directions $\hat{s}$ where the size is proportional to the top-hat ratio $N_s / \langle N_s^{\text{iso}} \rangle$, compared to the cosmic ray sources (star symbols). b) Reconstructed charges of the cosmic rays as a function of the true charges.}
\label{fig:GMF-signal}
\end{centering}
\end{figure}

For the astrophysical fit presented in Fig.~\ref{fig:GMF-fit}, we also investigate the reconstruction quality of the source directions and of the cosmic ray charges. To enhance signal cosmic rays, we use only cosmic rays with top-hat counts above isotropic expectations $N_s / \langle N_s^{\text{iso}} \rangle > 5$. In Fig.~\ref{fig:GMF-signal_a} we show the top-hat ratio of signal (background) cosmic rays using the colored (gray) symbols, whereby the symbol size follows the top-hat ratio. Also shown are the true source directions (star symbols). Regions of strong clustering are primarily found close to the sources.

Fig.~\ref{fig:GMF-signal_b} shows reconstructed cosmic ray charges in comparison to their true charges. While many of the charges of signal cosmic rays correlate with their true charges, background cosmic rays often receive too low or too high charges in order to join clusters.

\section{Conclusions}

In this paper we presented a new method for estimating the source directions of ultra-high energy cosmic rays by means of a simultaneous fit to all cosmic rays. Initially, the source directions and charges of the particles are estimated from measurements of their arrival direction on Earth, energy and shower depth. Assuming that some directions of origin are not isotropically distributed, but are instead concentrated in the direction of point sources, the estimated directions of origin and charges of the particles are iteratively improved within the fit. For this, we need information at least about the approximate local deflection directions of the galactic magnetic field, which we take from a parameterization of Faraday rotation and Starlight polarization measurements. The exact strength of the magnetic field is less important than the direction, because it can be compensated by a shift in the charge measurements.

The initially analyzed case of a 1-dimensional shift shows that the reconstruction of the sources is possible by the fit. The case also shows the challenge when signal particles and background particles have to be shifted in exactly the same direction and partially overlap. Here, half of the signal scenarios could still be distinguished from isotropic scenarios. In the 2-dimensional spherical case with deflections along the longitudinal coordinate, the fit allocates most particles correctly to their sources. After the fit, directions of particle origins cluster with a magnitude that does not occur in isotropic cases.

We tested an astrophysical scenario by simulating $N=1000$ different nuclei traversing the galactic magnetic field, where $m=4$ sources serve as sources of $10\%$ of the cosmic rays and $90\%$ of them follow an isotropic distribution. Here, too, we observe strong clustering of particle origins around the expected source directions. The clusters consist mainly of cosmic rays which were correctly allocated to their sources and, to a lesser extent, of particles of the isotropic admixture which are pushed by the fit towards the cluster. In purely isotropic scenarios, however, large clustering strength as observed in source scenarios is suppressed, such that source scenarios can be clearly separated from isotropic arrival distributions.

Overall, the fit method provides a novel approach for decoding the arrival distribution of ultra-high energy cosmic rays in terms of mixed-composition alignment patterns generated by deflections in the galactic magnetic field. It enables identification of cosmic ray directions outside our galaxy with a high occupancy and therefore candidate directions of cosmic ray point sources.

\section*{Acknowledgments}

We wish to thank Roger Clay for his valuable comments on the manuscript. This work is supported by the Ministry of Innovation, Science and Research of the State of North Rhine-Westphalia, and the Federal Ministry of Education and Research (BMBF).

\end{document}